\documentclass[10pt,conference]{IEEEtran}
\IEEEoverridecommandlockouts

\usepackage{xcolor}
\usepackage{booktabs} % For formal tables
\usepackage{tabularx}
\usepackage{colortbl}
\usepackage{pgf}
\usepackage{pgfplots}
\usepackage{etoolbox}
\usepackage{adjustbox}
\usepackage[linesnumbered,ruled,vlined]{algorithm2e}
\usepackage{amsmath}
 
\usepackage{amssymb}
\usepackage{bm}
\usepackage{enumitem}
\usepackage{threeparttable}  
\usepackage{multirow}
\usepackage{makecell}
\usepackage{tabularx}
\usepackage[T1]{fontenc}
\usepackage{array}
\usepackage{listings}
\usepackage{xspace} 
\usepackage{subcaption}
\usepackage{graphicx}
\usepackage{hyperref}
\usepackage{textcomp}
\usepackage{microtype} 
\usepackage{calc}
\usepackage{pifont}
\usepackage{tcolorbox}
\usepackage{tikz}
\usepackage{balance}

\usepackage{caption}
\newcolumntype{Y}{>{\centering\arraybackslash}X}

\SetCommentSty{mycommfont}
\SetKwInput{KwInput}{Input}                % Set the Input
\SetKwInput{KwOutput}{Output}              % set the Output

\definecolor{codegreen}{rgb}{0,0.6,0}
\definecolor{codegray}{rgb}{0.5,0.5,0.5}
\definecolor{codepurple}{rgb}{0.58,0,0.82}
\definecolor{backcolour}{rgb}{0.95,0.95,0.92}
\lstdefinestyle{mystyle}{
    backgroundcolor=\color{backcolour},   
    commentstyle=\color{codegreen},
    keywordstyle=\color{magenta},
    stringstyle=\color{codepurple},
    basicstyle=\ttfamily\footnotesize,
    breakatwhitespace=false,         
    breaklines=true, 
    postbreak=\mbox{\textcolor{gray}{$\hookrightarrow$}\space},
    captionpos=b,                    
    keepspaces=true,                 
    showspaces=false,                
    showstringspaces=false,
    showtabs=false,                  
    tabsize=2
}

\newlength{\ColorBoxDepthReference}
\newlength{\ColorBoxHeightReference}
\newlength{\Width}%
\newcommand{\MyColorBox}[2][red]%
{%
	\settowidth{\Width}{#2}%
	\colorbox{#1}%
	{%      
		\raisebox{-\ColorBoxDepthReference}%
		{%
			\parbox[b][\ColorBoxHeightReference+\ColorBoxDepthReference][c]{\Width}{\centering#2}%
		}%
	}%
}
\definecolor{winered}{RGB}{128, 0, 0}
\newcommand{\inlinecode}[1]{\textit{#1}}
\newcommand{\proc}[1]{\textbf{\textsc{#1}}}

\tcbuselibrary{skins}
\newcounter{finding}
\newcommand{\finding}[1]{\refstepcounter{finding}
	\begin{center}
		\begin{tcolorbox}[colback=gray!10,colframe=black!50,width=1\columnwidth,arc=1mm, auto outer arc,boxrule=0.5pt,boxsep=3pt,left=3pt,right=3pt,top=0pt,bottom=0pt]
		\textbf{Summary \arabic{finding}:} #1
		\end{tcolorbox}
	\end{center}
}

\newcommand{\researchquestion}[1]{
    \begin{tcolorbox}[enhanced,borderline west={2pt}{0pt}{gray!50},boxrule=0pt,frame empty,colback=white,sharp corners,width=\linewidth,boxsep=2pt,left=3pt,right=3pt,top=0pt,bottom=0pt]
    #1
    \end{tcolorbox}
}

\newcommand{\heatmapcell}[1]{%
    \pgfmathsetmacro{\value}{#1}%
    \ifdim \value pt < 2.5pt \cellcolor{orange!2.5}#1
    \else\ifdim \value pt < 5pt \cellcolor{orange!5}#1
    \else\ifdim \value pt < 7.5pt \cellcolor{orange!7.5}#1
    \else\ifdim \value pt < 10pt \cellcolor{orange!10}#1
    \else\ifdim \value pt < 12.5pt \cellcolor{orange!12.5}#1
    \else\ifdim \value pt < 15pt \cellcolor{orange!15}#1
    \else\ifdim \value pt < 17.5pt \cellcolor{orange!17.5}#1
    \else\ifdim \value pt < 20pt \cellcolor{orange!20}#1
    \else\ifdim \value pt < 22.5pt \cellcolor{orange!22.5}#1
    \else\ifdim \value pt < 25pt \cellcolor{orange!25}#1
    \else\ifdim \value pt < 27.5pt \cellcolor{orange!27.5}#1
    \else\ifdim \value pt < 30pt \cellcolor{orange!30}#1
    \else\ifdim \value pt < 32.5pt \cellcolor{orange!32.5}#1
    \else\ifdim \value pt < 35pt \cellcolor{orange!35}#1
    \else\ifdim \value pt < 37.5pt \cellcolor{orange!37.5}#1
    \else\ifdim \value pt < 40pt \cellcolor{orange!40}#1
    \else\ifdim \value pt < 42.5pt \cellcolor{orange!42.5}#1
    \else\ifdim \value pt < 45pt \cellcolor{orange!45}#1
    \else\ifdim \value pt < 47.5pt \cellcolor{orange!47.5}#1
    \else\ifdim \value pt < 50pt \cellcolor{orange!50}#1
    \else\ifdim \value pt < 52.5pt \cellcolor{orange!52.5}#1
    \else\ifdim \value pt < 55pt \cellcolor{orange!55}#1
    \else\ifdim \value pt < 57.5pt \cellcolor{orange!57.5}#1
    \else\ifdim \value pt < 60pt \cellcolor{orange!60}#1
    \else\ifdim \value pt < 62.5pt \cellcolor{orange!62.5}#1
    \else\ifdim \value pt < 65pt \cellcolor{orange!65}#1
    \else\ifdim \value pt < 67.5pt \cellcolor{orange!67.5}#1
    \else\ifdim \value pt < 70pt \cellcolor{orange!70}#1
    \else\ifdim \value pt < 72.5pt \cellcolor{orange!72.5}#1
    \else\ifdim \value pt < 75pt \cellcolor{orange!75}#1
    \else\ifdim \value pt < 77.5pt \cellcolor{orange!77.5}#1
    \else\ifdim \value pt < 80pt \cellcolor{orange!80}#1
    \else\ifdim \value pt < 82.5pt \cellcolor{orange!82.5}#1
    \else\ifdim \value pt < 85pt \cellcolor{orange!85}#1
    \else\ifdim \value pt < 87.5pt \cellcolor{orange!87.5}#1
    \else\ifdim \value pt < 90pt \cellcolor{orange!90}#1
    \else\ifdim \value pt < 92.5pt \cellcolor{orange!92.5}#1
    \else\ifdim \value pt < 95pt \cellcolor{orange!95}#1
    \else\ifdim \value pt < 97.5pt \cellcolor{orange!97.5}#1
    \else \cellcolor{orange}#1
    \fi\fi\fi\fi\fi\fi\fi\fi\fi\fi\fi\fi\fi\fi\fi\fi\fi\fi\fi\fi\fi\fi\fi\fi\fi\fi\fi\fi\fi\fi\fi\fi\fi\fi\fi\fi\fi
}

\newcommand{\mystar}{
    \textcolor[HTML]{8C564A}{\ding{72}}
}

\newcommand{\triaup}{
    \textcolor[HTML]{F5801C}{\tikz {
        \coordinate (A) at (0,0);
        \coordinate (B) at (-0.075,-0.15);
        \coordinate (C) at (0.075,-0.15);
        \coordinate (Center) at (0,-0.1);
        \draw[line width=0.8pt] (Center) -- (A);
        \draw[line width=0.8pt] (Center) -- (B);
        \draw[line width=0.8pt] (Center) -- (C);
    }}
}

\newcommand{\trialeft}{
    \textcolor[HTML]{2AA02D}{\tikz {
        \coordinate (A) at (0,0);
        \coordinate (B) at (0.15,-0.075);
        \coordinate (C) at (0.15,0.075);
        \coordinate (Center) at (0.1,0);
        \draw[line width=0.8pt] (Center) -- (A);
        \draw[line width=0.8pt] (Center) -- (B);
        \draw[line width=0.8pt] (Center) -- (C);
    }}
}

\newcommand{\eg}{\textit{e.g.,}\xspace}
\newcommand{\ie}{\textit{i.e.,}\xspace}
\newcommand{\etal}{\textit{et al.}\xspace}
\newcommand{\cf}{\textit{cf.}\xspace}

\newcommand{\mapping}[2]{\inlinecode{#1} $\to$ \inlinecode{#2}}
\newcommand{\buggy}{outdated\xspace}
\newcommand{\fixed}{up-to-dated\xspace}
\newcommand{\Buggy}{Outdated\xspace}
\newcommand{\Fixed}{Up-to-dated\xspace}
\newcommand{\good}{\emph{RepC}\xspace}
\newcommand{\bad}{\emph{DepC}\xspace}
\newcommand{\others}{\emph{Others}\xspace}
\newcommand{\plausible}{\emph{plausible}\xspace}

\newcommand{\aup}{AUP\xspace}
\newcommand{\dur}{DUR\xspace}

\newcommand{\wangchong}[1]{\textcolor{black}{#1}}
\newcommand{\revise}[1]{\textcolor{black}{#1}}

\newcommand{\hkf}[1]{#1}

\newcolumntype{C}[1]{>{\centering\arraybackslash}p{#1}}

\begin{document}
% \title[LLMs Meet Library Evolution: Evaluating Deprecated API Usage in LLM-based Code Completion]{LLMs Meet Library Evolution: \\Evaluating Deprecated API Usage in LLM-based Code Completion}
\title{LLMs Meet Library Evolution: \\Evaluating Deprecated API Usage in LLM-based Code Completion}

\author{
    \IEEEauthorblockN{Chong Wang\IEEEauthorrefmark{2}, Kaifeng Huang\IEEEauthorrefmark{3}\IEEEauthorrefmark{1}, Jian Zhang\IEEEauthorrefmark{2}, Yebo Feng\IEEEauthorrefmark{2}, Lyuye Zhang\IEEEauthorrefmark{2}, Yang Liu\IEEEauthorrefmark{2}, and Xin Peng\IEEEauthorrefmark{4}}
    \IEEEauthorblockA{
        \IEEEauthorrefmark{2}\textit{School of Computer Science and Engineering, Nanyang Technological University, Singapore}\\
        \{chong.wang, jian\_zhang, yebo.feng\}@ntu.edu.sg, zh0004ye@e.ntu.edu.sg, yangliu@ntu.edu.sg\\
        \IEEEauthorrefmark{3}\textit{School of Computer Science and Technology, Tongji University, China}\\
        kaifengh@tongji.edu.cn\\
        \IEEEauthorrefmark{4}\textit{School of Computer Science and Shanghai Key Laboratory of Data Science, Fudan University, China}\\ 
        pengxin@fudan.edu.cn
    }
    \thanks{\IEEEauthorrefmark{1} Kaifeng Huang is the corresponding author}
}

\maketitle

\begin{abstract}
    Large language models (LLMs), pre-trained or fine-tuned on large code corpora, have shown effectiveness in generating code completions. However, in LLM-based code completion, LLMs may struggle to use correct and up-to-date Application Programming Interfaces (APIs) due to the rapid and continuous evolution of libraries. While existing studies have highlighted issues with predicting incorrect APIs, the specific problem of deprecated API usage in LLM-based code completion has not been thoroughly investigated.

    To address this gap, we conducted the first evaluation study on deprecated API usage in LLM-based code completion. This study involved seven advanced LLMs, 145 API mappings from eight popular Python libraries, and 28,125 completion prompts. 
    % The study results reveal the \textit{status quo} and \textit{root causes} of deprecated API \hkf{and replacing API} usage in LLM-based code completion from the perspectives of \textit{model}, \textit{prompt}, and \textit{library}. 
    \hkf{The study results reveal the \textit{status quo} (\ie API usage plausibility and deprecated usage rate) of deprecated API \hkf{and replacing API} usage in LLM-based code completion from the perspectives of \textit{model}, \textit{prompt}, and \textit{library}, and indicate the \textit{root causes} behind.}
    Based on these findings, we propose two lightweight fixing approaches, \textsc{ReplaceAPI} and \textsc{InsertPrompt}, which can serve as baseline approaches for future research on mitigating deprecated API usage in LLM-based completion. Additionally, we provide implications for future research on integrating library evolution with LLM-driven software development.
\end{abstract}
\maketitle

\section{Introduction}
Large language models (LLMs) \cite{Codex,InCoder,CodeGen,StarCoder,CodeLlama,DeepSeek-Coder} have significantly advanced various aspects of software engineering~\cite{chen2025deep}, including code completion \cite{svyatkovskiy2020intellicode,le2022coderl,wang2024teaching}, code understanding \cite{zeng2022extensive,wang2024tiger}, defect detection~\cite{wang2023boosting,li2024llm}, and program repair \cite{fan2023automated,wei2023copiloting,zhang2024vuladvisor}. These models, pre-trained or fine-tuned with extensive knowledge of code on large corpora, are effective for tailoring to different downstream tasks. In the realm of code completion, the state-of-the-art has evolved from statistics-based methods \cite{nguyen2015graph, raychev2014code} to LLM-based techniques \cite{ugare2024improving, guo2022unixcoder, copilot}. Code completion is a sophisticated task that suggests variables, functions, classes, methods, and even entire code blocks, which depends on developers' practical needs. 
% Despite substantial advances, the full potential and risks of LLMs in code completion remains unclear.

% Code completion has long been a key focus in software engineering. Recent advancements have introduced a range of large language models (LLMs)~\cite{gpt-neo,gpt-j,Codex,CodeT5,CodeT5Plus,InCoder,AlphaCode,CodeGen,GPT-NeoX,SantaCoder,StarCoder,CodeLlama,PanGu-Coder}, based on the Transformer model architecture~\cite{Transformer}. These models can automatically complete code based on provided incomplete snippets by generating following statements. 

\textbf{Motivation.}
To accelerate development, developers heavily rely on third-party libraries, interacting with them through Application Programming Interfaces (APIs). However, this reliance presents a challenge for code completion tools. Third-party libraries constantly evolve to undergo refactorings \cite{Kula2018ESI}, fix bugs \cite{hu2023empirical}, apply security patches \cite{sigsoft/Wang0LC20}, or introduce new features. This rapid evolution leads to frequent API changes, with older APIs being deprecated and replaced by newer ones. Deprecated APIs are discouraged to use because of their incompatiblity with newer features or data, which will eventually disappear in future library updates \cite{APIlifecycle}. Taking PyTorch~\cite{PyTorch}, a popular deep learning library for instance, the API \inlinecode{torch.gels()} was deprecated in version 1.2 (August 2019) in favor of \inlinecode{torch.lstsq()}. Then, \inlinecode{torch.lstsq()} was deprecated in version 1.9 (June 2021) in favor of \inlinecode{torch.linalg.lstsq()}.~Consequently, newly developed code should avoid using the deprecated \inlinecode{torch.gels()} and \inlinecode{torch.lstsq()}. Therefore, it's crucial for code completion tools to suggest correct and up-to-dated APIs~to~developers.

% in code completion, the LLMs can select suitable Application Programming Interfaces (APIs) to achieve the desired functionality is understudied~\cite{ijcai/ZanCYLKGWCL22,corr/abs-2305-04032}.
\textbf{Literature.}
However, to the best of our knowledge, the capabilities of LLM-based code completion regarding API deprecation is understudied \cite{ijcai/ZanCYLKGWCL22,corr/abs-2305-04032}. Although there emerges a substantial number of evaluation on code completion, a body of research focused on assessing the overall accuracy across various benchmarks \cite{zeng2022extensive, xu2022systematic, liu2024your, ciniselli2021empiricaltransformer, ciniselli2021empiricalbert}. Interestingly, Ding \etal \cite{ding2023static} identified undefined names and unused variables as the most common syntactic errors produced by LLMs in Python code completions. Izadi \etal \cite{izadi2024language} found that incorrect function name predictions were prevalent, accounting for 23\% of all token-level errors. Furthermore, Recent studies \cite{liu2024exploring,zhang2024llm} highlighted the issue of hallucinations in LLM-generated code. Their findings indicate the prevalence and potential risks of using unexpected APIs. Nevertheless, while researchers have noted the prevalence of incorrect function name predictions, they have not investigated this issue in depth. Library APIs, which constitute an important part in predicting external function names, are worth attached importance to.

\textbf{Study.}
To address this gap, we conducted a study to examine the deprecated API usage in LLM-based code completion. The study aims to answer the primary research question: 
\researchquestion{\textbf{\textit{What are the status quo \hkf{(\ie API Usage Plausibility (AUP) and Deprecated Usage Rate (DUR))} and \revise{potential} root causes of deprecated and replacing API usage in LLM-based code completion?}}}
\noindent This question is explored through three detailed aspects: 

\noindent \textbf{Model Perspective (RQ1)} investigates the status quo and \revise{potential} causes based on the performance of various LLMs;

\noindent \textbf{Prompt Perspective (RQ2)} examines the impact of different prompts on the status quo and \revise{potential} causes; 

\noindent \textbf{Library Perspective (RQ3)} analyzes the status quo and \revise{potential} causes across different libraries.
% \begin{itemize}[leftmargin=15pt]
%     \item \textbf{Model Perspective (RQ1)} investigates the status quo and causes based on the performance of various LLMs;
%     \item \textbf{Prompt Perspective (RQ2)} examines the impact of different prompts on the status quo and causes;
%     \item \textbf{Library Perspective (RQ3)} analyzes the differences in status quo and causes across libraries.
% \end{itemize}

To address these research questions, we conducted a series of experiments involving various libraries and LLMs. We collected 145 API mappings between deprecated APIs and their replacements from eight popular Python libraries. Based on these mappings, we retrieved 9,022 \emph{\buggy} functions and 19,103 \emph{\fixed} functions using the deprecated APIs and replacing APIs, respectively. \hkf{Then, we identify the locating lines of the deprecated or replacing APIs, mask them and subsequent code, and use the remaining parts as the code completion prompts.}
% \wangchong{Each \buggy or \fixed function was transformed into a line-level completion prompt by identifying the deprecated or replacing API and removing the containing and subsequent lines.} 
\hkf{The original deprecated or replacing API is referred as the \emph{reference} API.} 
% \wangchong{The located API is referred to as the \emph{reference} API.} 
These prompts were then inputted into seven advanced code LLMs, including CodeLlama~\cite{CodeLlama} and GPT-3.5, to generate completions and analyze the predicted API usages. If the predicted API usage corresponds to either the deprecated or replacement version of the reference API, it is annotated as \plausible; otherwise, it is annotated as \others. 

The study results reveal the following findings:
\textbf{Finding in RQ1:} All evaluated LLMs encounter challenges in predicting \plausible API usages and face issues with deprecated API usages, due to the presence of deprecated API usages during model training and the absence of API deprecation knowledge during model inference.
\textbf{Finding in RQ2:} For the two categories of prompts derived from \buggy and~\fixed functions, the LLMs' performance in predicting \plausible and deprecated API usages differs significantly,~influenced~by the distinct code context characteristics of these prompts.
\textbf{Finding to RQ3:} Across the eight libraries, the LLMs~exhibit significant differences in their use of deprecated APIs, influenced by the characteristics of API deprecations during library~evolution. 

\textbf{Lightweight Mitigation.}
Based on the study results and findings, we \hkf{explored the feasibility of using} two lightweight fixing approaches to mitigate deprecated API usage in LLM-based code completion. Given a completion containing a deprecated API usage, the first approach, named \textbf{\textsc{ReplaceAPI}}, directly replaces the deprecated API usage with the replacement and regenerates the remaining parts (\eg argument list) during the decoding process. The second approach, named \textbf{\textsc{InsertPrompt}}, inserts an additional~replacing prompt after the original prompt to guide the LLMs to use the replacing API and then regenerate the completions.~We then evaluate the effectiveness of the proposed approaches in terms of fixing deprecated API usages and the accuracy in predicting line-level completions (\textbf{RQ4}).
The evaluation results demonstrate that \textsc{ReplaceAPI} effectively addresses deprecated API usages for all evaluated open-source LLMs, achieving fix rates exceeding 85\% with acceptable accuracy measured by Edit Similarity and Exact Match compared to ground-truth completions. While \textsc{InsertPrompt} does not currently achieve sufficient effectiveness and accuracy in fixing completions containing deprecated API usage, it shows potential for future~exploration.

\textbf{Contribution.} This paper makes the following contributions:
\begin{itemize}[leftmargin=15pt]
    \item We conducted the first study that reveals the status quo and causes of deprecated API usages in LLM-based code completion from prespectives of model, prompt, and library. 
    \item We proposed two lightweight approaches, named \textsc{ReplaceAPI} and \textsc{InsertPrompt}, to serve as baselines for mitigating deprecated API usage in LLM-based completion.
    \item We provide implications for future research on the synergy of library evolution and LLM-driven software~development.
\end{itemize}

% \vspace{-3mm}
\section{Related Work}

% We review the related work with respect to library evolution and LLM-based code completion.
\subsection{Library Evolution}
 
% deprecated api很广泛
Library evolution involves refactorings \cite{Kula2018ESI}, bug fixes \cite{hu2023empirical}, and new feature introductions. Typically, refactorings can deprecate old APIs and introduce new replacements. Several studies have examined the reasons that developers deprecate APIs and how the clients react to such deprecations~\cite{sawant2018understanding, sawant2018features, mirian2019web, sawant2019react}. The reasons include improving readability, reducing redundancy, avoiding bad code practices and fixing functional bugs. Deprecated APIs can affect hundreds of clients \cite{robbes2012developers}, particularly when clients struggle to keep pace with rapidly evolving software \cite{Linares-Vasquez2013ACF, wang2020empirical}. McDonnell et al. \cite{McDonnell2013ESA} found that only 22\% of outdated API usages are eventually upgraded to use replacing APIs. Similarly, Hora et al. \cite{hora2015developers} found that client developers consumes considerate time to discover and apply replacing APIs, with the majority of systems not reacting at all. When clients do not upgrade their APIs, they silently accumulate technical debt in the form of future API changes when they finally upgrade \cite{sawant2016reaction}. 
% \todo{To enjoy Therefore, in the context of code completion, it is better not to use deprecated API.}
% deprecated api
To locate the replacing API, existing works leverage change rules written by developers \cite{Balaban2005RSC}, developer recordings \cite{henkel2005catchup}, similarity matching \cite{xing2007api}, mining API usage in libraries \cite{dagenais2009semdiff}, and in client projects \cite{schafer2008mining}. Henkel and Diwan \cite{henkel2005catchup} developed an IDE plugin that allows library developers to record API refactoring actions and client developers to replay them. Godfrey and~Zou \cite{godfrey2005using} proposed a semi-automated origin analysis using similarities in name, declaration, complexity metrics, and call dependencies. Wu et al. \cite{wu2010aura} introduced a hybrid approach combining call dependency and text similarity analysis to identify API change rules. Recently, \cite{Huang2021} proposed RepFinder to find replacing APIs for deprecated APIs in library updates from multiple~sources.
% \todo{Our work}

In this work, we aim to comprehend the \wangchong{statuses and causes} of deprecated API usage in LLM-based code completion and provide implications for mitigating the deprecated API usages.

% To adapt the replacing API in the original context in client project, a simple approach is to replace calls to a deprecated API with its body \cite{perkins2005automatically}. Several works rely on historical update examples~\cite{nguyen2010graph, xu2019meditor, fazzini2019automated, thung2020automated, lamothe2020a3}. Haryono et al. \cite{haryono2020automatic} proposed AppEvolve, which relies on having before- and after-update examples to conduct usage updates for deprecated APIs. Thung et al. \cite{thung2019towards} proposed using a combination of API source code analysis and generated transformation rules to assist developers in replacing deprecated APIs without examples. Gao et al. \cite{gao2021apifix} proposed output-oriented program synthesis to automate API usage adaptations via program transformation. \todo{Our work}

\subsection{LLM-based Code Completion}

Code completion is an important functionality in modern IDEs and editors. Historically, researchers have explored statistical models \cite{nguyen2015graph, raychev2014code}. With the advent in natural language processing, researchers have embraced deep learning for code completion \cite{svyatkovskiy2020intellicode, le2022coderl} because they are similar in token-based prediction. 
To explore the capability of code completion tools driven by LLMs \cite{ugare2024improving, guo2022unixcoder, copilot}, numerous evaluations of LLMs have been proposed. Ciniselli et al. \cite{ciniselli2021empiricaltransformer, ciniselli2021empiricalbert} conducted a large-scale study exploring the accuracies of state-of-the-art Transformer-based models in supporting code completion at various granularity levels, from single tokens to entire code segments. Zeng et al. \cite{zeng2022extensive} found that pre-trained models significantly outperform non-pre-trained state-of-the-art techniques in program understanding tasks. They also reveal that no single pre-trained model dominates across all tasks. Xu et al. \cite{xu2022systematic} evaluated the performance of LLMs on the HumanEval dataset. Ding et al. \cite{ding2023static} identified undefined names and unused variables as the most common errors produced by language models in Python code completions. Izadi et al. \cite{izadi2024language} evaluated the LLMs using real auto-completion usage data across 12 languages. They found that incorrect function name predictions, were prevalent, accounting for 23\% of all token-level errors. Besides, Liu et al. \cite{liu2024your} proposed EvaluPlus, which benchmarks the functional correctness of LLM-synthesized code using test cases. In addition to accuracy concerns, LLM-based approaches face issues such as security vulnerabilities and hallucinations. Sallou et al. \cite{sallou2024breaking} explored threats posed by LLMs, including unpredictability in model evolution, data leakage, and reproducibility. Liu et al. \cite{liu2024exploring} categorized the hallucinations brought by LLM-generated code. 

The findings on incorrect function predictions partially motivate our study. However, we focus on the severity of predicting deprecated API usages in LLM-based code completion.

\section{Study Setup}

\begin{figure}
    \centering
    \includegraphics[width=1\columnwidth]{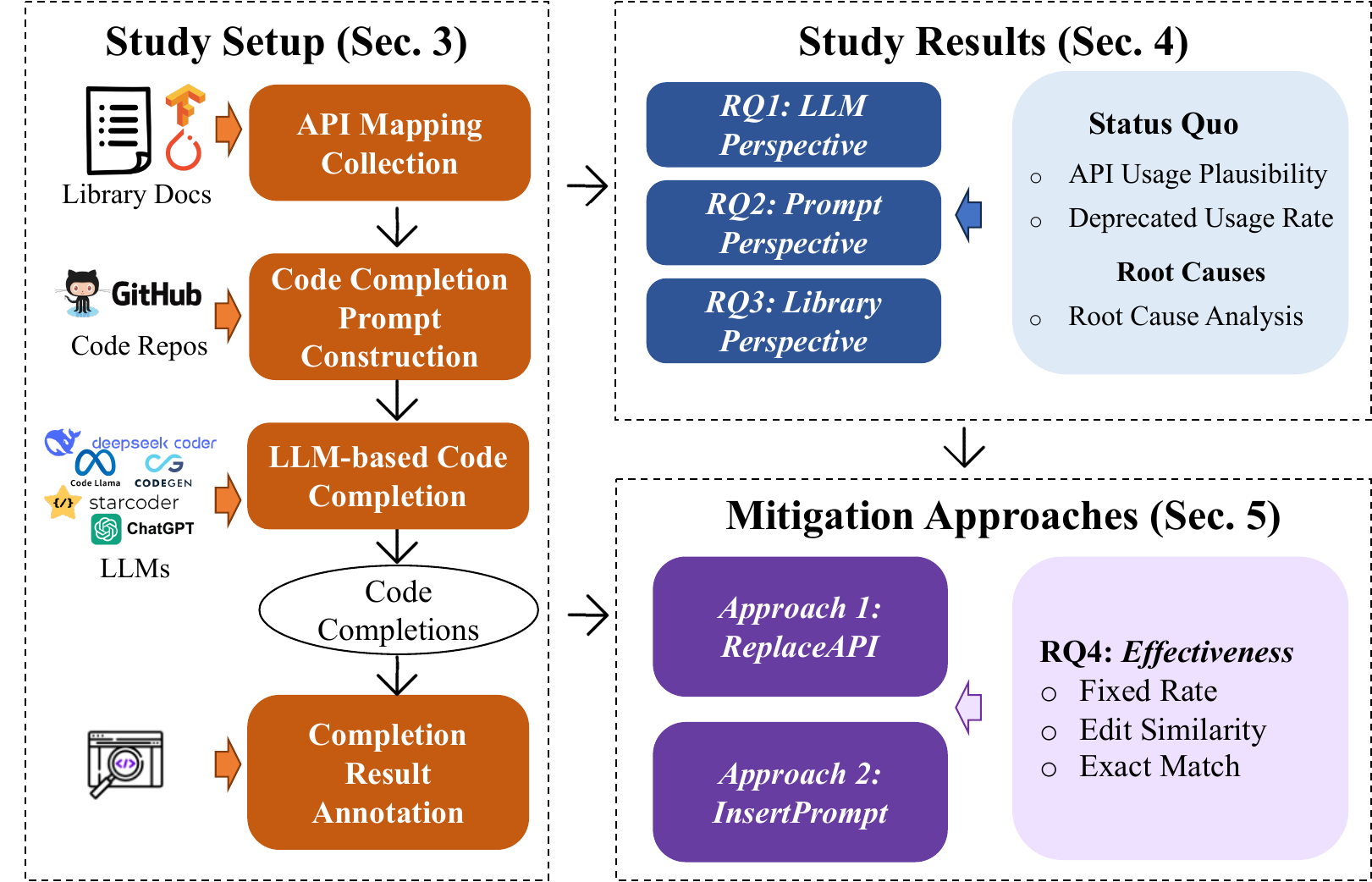}
    \vspace{-14pt}
    \caption{Overview of Our Study}
    \label{fig:setup-overview}
\end{figure}

% hkf:
We chose Python, a popular programming language which ranks first among the most popular ones based in the recent year \cite{popularlanguage}. We targeted eight popular Python libraries. 
% Five of these libraries were used in a previous study on Python API deprecation~\cite{sigsoft/Wang0LC20}, including Numpy~\cite{Numpy}, Pandas~\cite{Pandas}, scikit-learn~\cite{scikit-learn}, SciPy~\cite{SciPy}, and seaborn~\cite{seaborn}. Additionally, we added three popular deep learning libraries, \ie TensorFlow~\cite{TensorFlow}, PyTorch~\cite{PyTorch}, and Transformers~\cite{Transformers}. 
Five of these libraries were used in a previous study on Python API deprecation~\cite{sigsoft/Wang0LC20}, including Numpy, Pandas, scikit-learn, SciPy, and seaborn. Additionally, we added three popular deep learning libraries, \ie \href{https://www.tensorflow.org/}{TensorFlow}, \href{https://pytorch.org/}{PyTorch}, and \href{https://huggingface.co/docs/transformers/index}{Transformers}. 
The setup of our study is presented in Figure \ref{fig:setup-overview}. It includes four steps. The \textit{API Mapping Collection} gathers mappings between deprecated APIs and their replacements from various libraries. The \textit{Completion Prompt Construction} step involves creating completion prompts by identifying instances of deprecated and replacing API usage in open-source Python repositories. The \textit{LLM-Based Code Completion} step uses various LLMs to generate code completions for these prompts. Finally, the \textit{Completion Result Annotation} step automatically annotates the generated completions and calculates relevant metrics.

\subsection{API Mapping Collection}

% We selected 8 popular Python libraries to collect deprecated APIs and their replacements. Five of these libraries were also used in a previous study on Python API deprecation~\cite{sigsoft/Wang0LC20}, including Numpy~\cite{Numpy}, Pandas~\cite{Pandas}, scikit-learn~\cite{scikit-learn}, SciPy~\cite{SciPy}, and seaborn~\cite{seaborn}. The remaining 3 are newer deep learning libraries: TensorFlow~\cite{TensorFlow}, PyTorch~\cite{PyTorch}, and Transformers~\cite{Transformers}
% hkf:
We identified API mappings (\ie deprecated APIs and the mapping replacements) from the documentation and change logs from each library version following the previous study~\cite{sigsoft/Wang0LC20}. \hkf{For each library, we selected their versions that were released after January 2019. 
\revise{We chose January 2019 as the starting point for collecting API mappings, providing a five-year window that balances sufficient version numbers with manageable human effort.}
The version ranges are presented in Table~\ref{tab:dataset}.} Specifically, we reviewed the documentation and change logs of each library and manually look for deprecated API occurrences which indicate the corresponding the mapping replacements. For instance, in the API documentation of PyTorch, version 1.9.0 \cite{pytorchdoc}, a deprecation message indicates that \emph{``torch.lstsq() is deprecated in favor of torch.linalg.lstsq() and will be removed in a future PyTorch release.''}, where the mapping of the deprecated API to the replacing API is \mapping{torch.lstsq}{torch.linalg.lstsq}. For one-to-many mappings  (\ie one deprecated API mapped to multiple replacing APIs), we split them into many one-to-one mappings. \revise{Two authors independently collected the data, with a third author resolving inconsistencies. The process took about three working days, and the Jaccard coefficient between the first two collectors' mappings was 91.8\%.} \hkf{In total, we collected a preliminary number of 247 API mappings.}

%In total, we obtained 145 API mappings. 

%The statistics of our selected library versions and API mappings are presented Table~\ref{tab:dataset}.

% For each library, we manually reviewed its documentation and change logs to identify API deprecations and collect mappings between deprecated APIs and their replacements. For instance, in the version 1.9.0 PyTorch API documentation, a deprecation warning was highlighted: \emph{``torch.lstsq() is deprecated in favor of torch.linalg.lstsq() and will be removed in a future PyTorch release.''} From this deprecation warning, we collected the mapping of \mapping{torch.lstsq}{torch.linalg.lstsq}. In total, we obtained 145 API mappings for all the 8 libraries. Detailed statistics, grouped by library, are presented in Table~\ref{tab:dataset}. 

\begin{table}
    % \caption{Dataset statistics.}
    \caption{Statistics of our Collected API Mappings}
    \vspace{-6pt}
    \label{tab:dataset}
    \centering
    \renewcommand{\arraystretch}{1.2}
    \setlength{\tabcolsep}{5pt}
    \begin{tabularx}{1.0\columnwidth}{l|cccc} 
    \Xhline{2\arrayrulewidth}
    \multirow{2}{*}{\textbf{Library}} & \multirow{2}{*}{\textbf{Version Range}} & \multirow{2}{*}{\textbf{\# Mappings}} & \multicolumn{2}{c}{\textbf{\# Functions}} \\
    \cline{4-5}
                                      &                                   &                                       &  \Buggy    &     \Fixed    \\ 
    % \cline{1-1}\cline{3-3}\cline{5-7}\cline{9-10}
    \Xhline{1.5\arrayrulewidth}                           
    Numpy                             &           1.16.0-1.26.4           &           3                           &   567      &     2,988     \\                          
    Pandas                            &           0.24.0-2.2.2            &          10                           &    69      &       69      \\                         
    scikit-learn                      &           0.21.3-1.5.0            &          18                           &   985      &     1,197     \\                    
    SciPy                             &           1.2.1-1.13.0            &           4                           &   245      &      1,458    \\                        
    seaborn                           &           0.9.1-0.13.2            &           3                           &   904      &      1,329    \\                         
    TensorFlow                        &           1.13.1-2.16.1           &          57                           &  1,491     &      4,830    \\                         
    PyTorch                           &           1.0.1-2.3.0             &          21                           &  4,726     &     6,406     \\                         
    Transformers                      &           1.0.0-4.40.2                  &          29                           &   100      &       63      \\
    \Xhline{1.5\arrayrulewidth}     
    Total                             &           --                      &         145                           &  9,022     &     19,103    \\
    \Xhline{2\arrayrulewidth}
    \end{tabularx}
\end{table}

\subsection{Completion Prompt Construction}\label{sec:matching}
% hkf:
We constructed code completion prompts by searching \revise{real-world} code snippets that contain usages of either deprecated APIs or replacing APIs from collected API mappings in open-source~Python~repositories, \revise{to simulate realistic code completion scenarios}.
% the deprecated API and replacing API usages from . 

\subsubsection{\Buggy and \Fixed Function Location.}\label{sec:setup:ccpc:matching}
We utilized Sourcegraph~\cite{Sourcegraph}, a widely used code search service \hkf{to search code snippets}. It supports integration with GitHub where we can retrieve Python source files from millions of open-source code repositories. For each deprecated or replacing API, we constructed search queries using both its full qualified name (FQN) (\eg \inlinecode{torch.lstsq}) and a logical disjunction of its constituent parts (\eg ``torch AND linalg AND lstsq'') to ensure comprehensive retrieval. For each retrieved Python source file, we parsed it into an Abstract Syntax Tree (AST) and extracted the containing functions that invoked the deprecated or replacing APIs. Specifically, we located function definition nodes in the AST and traversed its descendants. For~each~descendant, we checked if it is a function call node and matched the function call to the deprecated or replacing APIs. To correctly match the function call via~API~FQNs, we performed lightweight object type~resolution and alias resolution, similar~to~\cite{sigsoft/Wang0LC20}.

% During this process, we performed lightweight object type resolution and API alias resolution, similar to~\cite{sigsoft/Wang0LC20}, to normalize the calls into their corresponding API FQNs.
\begin{itemize}[leftmargin=*]
    \item \textbf{Object Type Resolution}: 
    In the object-oriented programming (OOP) languages, the APIs can be encapsulated into a class as a method. Therefore, determining the FQN of the API invocation need to resolve the corresponding type of the invoking object. For example, the pandas library defines a core class \inlinecode{DataFrame} with a member method \inlinecode{loc()} and the client creates an object of class \inlinecode{DataFrame}, assigns to a variable \inlinecode{dt}, and invokes the method using \inlinecode{dt.loc()}. Typically, it requires resolving the type of \inlinecode{dt}. To that end, we analyzed the \texttt{assign} statements to track object definitions, enabling us to determine the class names for objects in function calls and infer the called APIs. For instance, if the object ``dt'' in the call \inlinecode{dt.loc()} was created in a preceding \texttt{assign} statement (\inlinecode{dt = pandas.DataFrame(...)}), we could infer that the corresponding API was \inlinecode{pandas.DataFrame.loc()}.
    % \item \textbf{Object Type Resolution}: In Python, object-oriented programming (OOP) allows APIs to be provided as classes and class methods. For example, the pandas library defines a core class \inlinecode{DataFrame} with a member method \inlinecode{DataFrame.loc()}. To handle calls to such APIs, we analyzed \texttt{assign} statements to track object definitions, enabling us to determine the class names for objects in function calls and infer the called APIs. For instance, if the object ``dt'' in the call \inlinecode{dt.loc()} was created in a preceding \texttt{assign} statement (\inlinecode{dt = pd.DataFrame(...)}), we could infer that the corresponding API was \inlinecode{pandas.DataFrame.loc()}.
    \item \textbf{Alias Resolution}: Developers can alias packages, classes, and functions in Python using the \texttt{import-as} feature~\cite{PEP-221}. This mechanism requires resolving API aliases by analyzing \texttt{import} statements. For example, the pandas package~is often imported with the alias ``pd'' via the statement \inlinecode{import pandas as pd}. In this case, \inlinecode{pd.DataFrame.loc()} was resolved to \inlinecode{pandas.DataFrame.loc()}. Additionally, Python provides a \texttt{from-import} mechanism allowing developers to use APIs with short names instead of their FQNs. For example, through \inlinecode{from torch.linalg import lstsq}, the API in \inlinecode{torch.linalg} can be directly called via \inlinecode{lstsq()}. These short names were resolved by analyzing the \texttt{import}~statements.
    % \item \textbf{API Alias Resolution}: Developers can alias packages, classes, and functions in Python using the \texttt{import-as} feature~\cite{PEP-221}. This aliasing mechanism necessitates resolving API aliases by analyzing \texttt{import} statements. For example, the pandas package is often imported with the alias ``pd'' via the statement \inlinecode{import pandas as pd}. In this case, \inlinecode{pd.DataFrame.loc()} was resolved to \inlinecode{pandas.DataFrame.loc()}. Additionally, Python provides a \texttt{from-import} mechanism allowing developers to use APIs with short names instead of their FQNs. For example, through \inlinecode{from torch.linalg import lstsq}, \inlinecode{torch.linalg.lstsq()} can be directly called as \inlinecode{lstsq()}. These short names were also resolved by analyzing the \texttt{import} statements.
\end{itemize}

After the lightweight object type resolution and alias resolution, we obtained the corresponding FQN for each function call. We checked whether the corresponding FQNs matched the APIs in the collected API mappings, identifying the first matched API as the \emph{reference} API. 
% After normalizing the function calls, we checked whether the corresponding FQNs matched the previously identified deprecated APIs and replacement APIs. As illustrated in Figure~\ref{fig:prompt}, the \fixed function calls \inlinecode{torch.linalg.solve()}, which is the replacement of the deprecated API \inlinecode{torch.solve()}.

We denote the containing function as an \emph{\buggy} function if a deprecated API was matched. Meanwhile, we denote it as an \emph{\fixed} function if a replacing API was matched. We collected 113,660 Python source files by querying SourceGraph. \hkf{We filtered out API mappings that returned either no instances of \emph{\buggy} or \emph{\fixed} functions. As a result, we collected 9,022 \emph{\buggy} and 19,103 \emph{\fixed} functions from 145 API mappings.} The statistics are presented in Table~\ref{tab:dataset}.

% If the FQN of the function call matches that of the replacement API, 
% If a function's reference API was deprecated, the function was labeled as \emph{\buggy}; if it was a replacement API, it was labeled as \emph{\fixed}. 
% Totally, we obtained 9,022 \buggy and 19,103 \fixed functions, respectively. 

% Functions matching multiple calls of deprecated and replacement APIs retained only the first ones, denoted as the \emph{reference} APIs. If a function's reference API was deprecated, the function was labeled as \emph{\buggy}; if it was a replacement API, it was labeled as \emph{\fixed}. Totally, we obtained 9,022 \buggy and 19,103 \fixed functions, respectively. Detailed statistics, grouped by library, are presented in Table~\ref{tab:dataset}.

% Note that, for Pandas, the numbers of extracted \buggy and \fixed functions were small, as the type resolution for the calls of Pandas APIs is usually difficult. For instance, the objects of \inlinecode{pandas.Series} are often created by performing slicing operation on \inlinecode{pandas.DataFrame} like \inlinecode{s = df[:]}. 

\begin{figure}
    \centering
    \includegraphics[width=1.0\columnwidth]{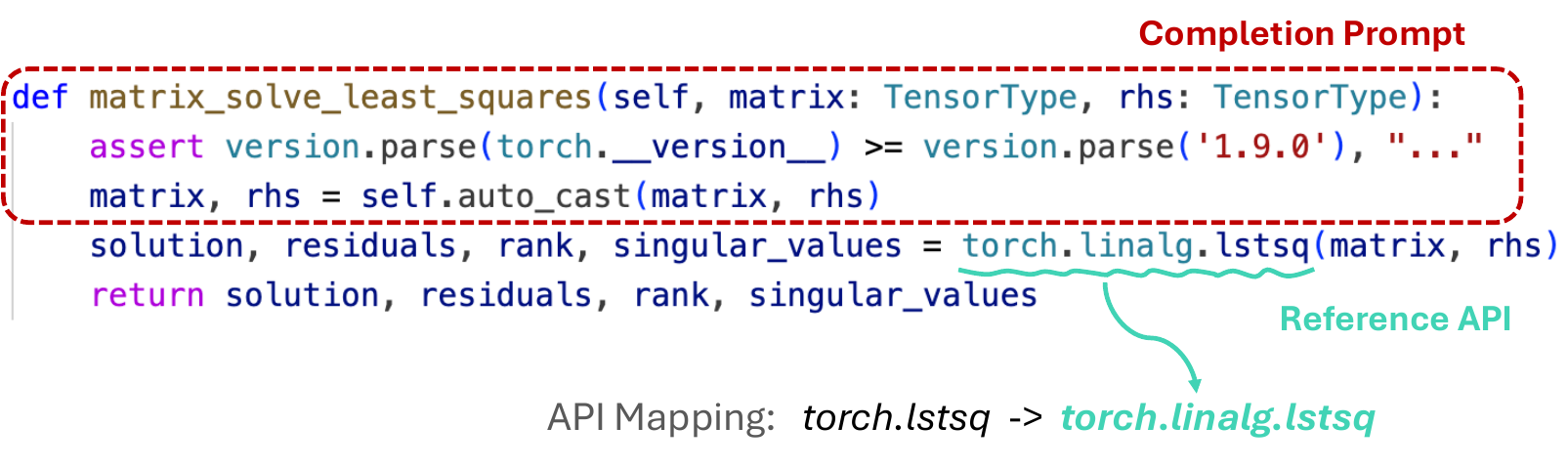}
    \vspace{-18pt}
    \caption{Illustration of Completion Prompt Construction for An \Fixed Function.}
    \label{fig:prompt}
\end{figure}

\subsubsection{Incomplete Code Extraction.}

In the task of code completion, developers usually have started with a few lines of code and pause in the middle, waiting for LLMs to return the suggested content based on the upward context. Therefore, to evaluate the performance of LLMs in the scenario, we constructed the \emph{line-level code completion prompts}. For each \buggy or \fixed function, we located the invocation line of the deprecated or replacing APIs, respectively. We collected the preceding lines before the invocation line into our \emph{line-level code completion prompts} for an \buggy or \fixed function, which is usually incomplete.

Figure~\ref{fig:prompt} represents one of our collected \fixed functions. The function invokes a API of PyTorch, \ie \inlinecode{torch.linalg.lstsq} in the fourth line. The line-level code completion prompt for this function is highlighted in the \textcolor{winered}{wine-red} dotted rectangle.

% he preceding lines, , are used as a line-level code completion prompt.

After processing all \buggy and \fixed functions, we obtained two corresponding datasets, denoted as $\mathcal{O}$ and $\mathcal{U}$, respectively. Each sample in $\mathcal{O}$ and $\mathcal{U}$ was formatted as ($pmpt$, $dep \to rep$), where $pmpt$ is a code completion prompt $pmpt$ and $dep \to rep$ denotes an API mapping from the deprecated API to the corresponding replacing API. 
% \hkf{Fig. \ref{fig:rq-box} presents the boxplots of $pmpt$'s LOC in $\mathcal{O}$ and $\mathcal{U}$, indicating that the length of $pmpt$ w.r.t lines of code (LOC) are similar in both datasets. Besides, the median LOC of $pmpt$ is 11 in both $\mathcal{O}$ and $\mathcal{U}$, which provides sufficient context for the code completion task.}
% \hkf{TODO5 How many lines are fed into LLM for code completion? Are those line contexts enough and sufficient for a code completion LLM task? }

% After processing all \buggy and \fixed functions, we obtained two corresponding datasets, denoted as $\mathcal{B}$ and $\mathcal{F}$, respectively. Each sample in $\mathcal{B}$ and $\mathcal{F}$ was formatted as ($pmpt$, $dep \to rep$), consisting of a code completion prompt $pmpt$ and a mapping between deprecated API and the replacement $dep \to rep$.

% \begin{figure}
%     \includegraphics[width=0.7\columnwidth]{img/loc.pdf}
%     \centering
%     \vspace{-4pt}
%     \caption{Boxplots of $pmpt$'s LOC in $\mathcal{O}$ and $\mathcal{U}$}
%     \label{fig:rq-box}
%     \vspace{-14pt}
% \end{figure}

\subsection{LLM-based Code Completion}

%We leverage multiple LLMs in the code completion task to observe their performance. The LLMs include both open-source and closed-source models, whose parameter sizes ranges from 350 million to 175 billion. The complete list of the LLMs are presented in Table~\ref{tab:llms}.

We leverage multiple LLMs including open-source and closed-soure with varying parameter sizes and observe their performance on the code completion task. The complete LLM list is presented in Table~\ref{tab:llms}.

\begin{itemize}[leftmargin=12pt]
    \item \textbf{CodeGen-350m, 2b, 6b}: CodeGen~\cite{CodeGen} is a family of LLMs developed by Salesforce specifically for code generation. 
    % These models are trained on diverse programming languages and are designed to assist in writing code by providing intelligent code suggestions and completions. 
    \item \textbf{DeepSeek-1.3b}: DeepSeek-Coder~\cite{DeepSeek-Coder} is designed on top of Transformer and tailored for code-related applications. 
    % This model is trained on a diverse set of coding repositories, which include a variety of programming languages and coding styles. It combines state-of-the-art machine learning techniques to provide robust code suggestions and completions.
    \item \textbf{StarCoder2-3b}: StarCoder2~\cite{StarCoder} is an LLM optimized for coding tasks. It leverages large-scale pre-training on code datasets to understand programming languages deeply. 
    % StarCoder2 excels in code completion, bug detection, and code translation, making it a valuable tool for developers seeking to enhance their coding efficiency and accuracy.
    \item \textbf{CodeLlama-7b}: CodeLlama~\cite{CodeLlama} is a specialized variant of Meta's LLaMA~\cite{Llama2}, adapted for programming tasks. 
    % It is designed to support developers by providing code suggestions, completing code snippets, and assisting in debugging. CodeLlama is trained on a vast corpus of programming languages and can perform a variety of code-related tasks with high proficiency.
    
    \item \textbf{GPT-3.5}: GPT-3.5~\cite{GPT-3.5} is a general-purpose language model developed by OpenAI. While it is not exclusively designed for coding, it possesses powerful code generation capabilities due to its extensive training on diverse text, including programming languages. 
    % GPT-3.5 can understand and generate code across various languages, making it a versatile tool for code completion, documentation, and coding assistance. When used for code-related tasks, an instruction (prompt) is often provided to guide the model to generate the desired code output.
\end{itemize}

The first six code LLMs were downloaded from Hugging Face~\cite{HuggingFace}. For LLMs with Python-specific versions available, we utilized those versions to improve completion results for Python functions. Consequently, the versions used for CodeGen and CodeLlama were \textit{CodeGen-\{350M,2B,6B\}-mono}~and \textit{CodeLlama-7b-Python-hf}, which were~fine-tuned~on~additional Python corpora. For DeepSeek and StarCoder2, the versions employed were \textit{deepseek-coder-1.3b-instruct},~\textit{starcoder2-3b}, respectively. For the general purpose LLM, \ie GPT-3.5, we queried the model via its official online APIs~\cite{OpenAIAPI}, using the \textit{gpt-3.5-turbo} version released in January 2024.

% For the samples in $\mathcal{O}$ and $\mathcal{U}$, we fed the prompts into the LLMs  to perform line-level code completion. 
% To feed the prompts from $\mathcal{O}$ and $\mathcal{U}$ into the LLMs to perform

For the six LLMs specifically tailored for code-related tasks, \ie three versions of CodeGen, DeepSeek, StarCoder2, and CodeLlama, the constructed prompts can be directly fed into the models to generate completions. Meanwhile, for the general-purpose GPT-3.5, we provided an instruction \revise{to specify the task, preventing the model from performing beyond code completion.} The instruction was: \emph{``Complete and output the next line for the following Python function: {$pmpt$}''}. For all these LLMs, we utilized greedy decoding (\ie choosing the token with highest possibility at each decoding step) to generate one completion for each prompt in $\mathcal{O}$ or $\mathcal{U}$. The maximal output token limit was set to 50. The greedy decoding for GPT-3.5 was implemented through setting the temperature parameter to 0. 
% The first six LLMs (\ie three versions of CodeGen, DeepSeek, StarCoder2, and CodeLlama) are tailored specifically for code-related tasks, allowing the constructed prompts to be directly input for generating completions. 

% To implement greedy search with GPT-3.5, we set the temperature parameter to 0 when querying the model~\cite{OpenAIAPI}. 

The procedure of LLM-based completion is defined as: $$ comp \gets \textbf{LLM}(pmpt) $$
% The first six LLMs (\ie three versions of CodeGen, DeepSeek, StarCoder2, and CodeLlama) are tailored specifically for code-related tasks, allowing the constructed prompts to be directly input for generating completions. However, for the general-purpose GPT-3.5, we provided an instruction to ensure accurate code completion for the constructed prompts. The instruction used was: \emph{``Complete and output the next line for the following Python function: {$pmpt$}''}. For all these LLMs, we utilized greedy search to generate one completion for each prompt, limiting the output to a maximum of 50 tokens. To implement greedy search with GPT-3.5, we set the temperature parameter to 0 when querying the model~\cite{OpenAIAPI}. Formally, the procedure of LLM-based completion is defined as: $$ comp = \textbf{LLM}(pmpt) $$

\begin{table}[!t]
    \caption{Evaluated Large Language Models (LLMs)}
    \label{tab:llms}
    \centering
    \renewcommand{\arraystretch}{1.2}
    \setlength{\tabcolsep}{4pt}
    \begin{tabular}{l|r|c|c} 
    \Xhline{2\arrayrulewidth}
    \textbf{Model} &  \textbf{\# Params} & \revise{\textbf{Python Fine-Tuned}}  & \textbf{Open-Source}  \\
    \Xhline{1.5\arrayrulewidth}          
    CodeGen-350m    &  350 M  &  \ding{51}  & \ding{51}  \\
    CodeGen-2b      &  2 B    &  \ding{51}  & \ding{51}  \\
    CodeGen-6b      &  6 B    &  \ding{51}  & \ding{51}  \\
    DeepSeek-1.3b   &  1.3 B  &  \ding{55}  & \ding{51} \\
    StarCoder2-3b   &  3 B    &  \ding{55}  & \ding{51}    \\
    CodeLlama-7b    &  7 B    &  \ding{51}  & \ding{51}   \\
    GPT-3.5         &  175 B  &  \textbf{UNKNOWN}  & \ding{55}      \\ 
    \Xhline{2\arrayrulewidth}
    \end{tabular}
    
    % \begin{tabular}{m{1.3cm}m{1cm}m{0.7cm}||m{1.3cm}m{1cm}m{0.7cm}} 
    % \Xhline{2\arrayrulewidth}
    % \textbf{Model} &  \textbf{\# Params}  & \textbf{Open-Source} &  \textbf{Model}  & \textbf{\# Params}  & \textbf{Open-Source}  \\
    % % \Xhline{1.5\arrayrulewidth}          
    % CodeGen-350m  &  350 M    & \ding{51}  & StarCoder2-3b   &   3 B    &     \ding{51}    \\ \Xhline{1\arrayrulewidth}
    % CodeGen-2b    &  2 B      & \ding{51}  &  CodeLlama-7b  &  7 B        &     \ding{51}   \\ \Xhline{1\arrayrulewidth}
    % CodeGen-6b    &    6 B    &  \ding{51}    &  GPT-3.5   &      175 B   &   \ding{55}      \\ 
    % \Xhline{1\arrayrulewidth}
    % DeepSeek-1.3b    &     1.3 B          &     \ding{51}   & & &  \\ 
    % \Xhline{2\arrayrulewidth}
    % \end{tabular}

\end{table}

    %        \\ 
    % \Xhline{1\arrayrulewidth}
    %        \\ 
    % \Xhline{1\arrayrulewidth}
    %        \\ 
    % \Xhline{2\arrayrulewidth}

\subsection{Completion Result Annotation}

For each sample ($pmpt$, $dep \to rep$), we examined the completions generated by LLMs and determine whether the studied API was predicted and whether the deprecated API or replacing API was predicted. Specifically, we extract the FQN of the API invocation in the predicted line using the same object type resolution and alias resolution in Sec. \ref{sec:setup:ccpc:matching}. The annotation procedure is formally described as follows:%, $comp$ is the completion result:
\hkf{$$ \{ \bad, \good, \others \} \leftarrow \text{anno}(comp) $$}
\hkf{Specifically, \bad denotes that the LLM gives a Deprecated Completion suggestion, and \good denotes that the LLM gives a Replacing Completion suggestion.}
\hkf{We identify \bad and \good by matching the FQN of an invocating API to either $dep$ or $rep$.}
%\ie the FQN of an invocating API was matched to $dep$;  \ie the FQN of an invocating API was matched to $rep$; 
\others denotes that the LLM suggests neither of the mapping APIs. Moreover, if a completion was annotated as either \bad or \good, we treat it as \plausible. This indicates that the LLM successfully understood the code context and selected a plausible API functionality.

\subsection{Metrics}\label{sec:metrics}
% Based on the annotated result, 
We investigate the performance of the LLMs using the following metrics: 
\begin{itemize}[leftmargin=15pt]
    \item \noindent\textbf{API Usage Plausibility (\aup):} This metric measures the portion of \plausible completions, which were annotated as \bad or \good. \aup is defined as:

\noindent$$ \aup = \frac{1}{|\mathcal{P}|} \sum_{p \in \mathcal{P}} \mathbb{I}(\text{anno}(\textbf{LLM}(p)) \in \{\bad, \good \}) $$
    \item \noindent\textbf{Deprecated Usage Rate (\dur):} This metric calculates the rate of \plausible completions that were annotated as \bad. \dur is defined as:

\noindent$$ \dur = \frac{\sum_{p \in \mathcal{P}} \mathbb{I}(\text{anno}(\textbf{LLM}(p)) = \bad)}{\sum_{p \in \mathcal{P}} \mathbb{I}(\text{anno}(\textbf{LLM}(p)) \in \{ \bad, \good\})} $$
\end{itemize}

\noindent$\mathcal{P}$ is the prompt set (\ie $\mathcal{O}$ or $\mathcal{U}$), and $\mathbb{I}(\cdot)$ is a binary function that returns 1 if the passed argument is true and 0 otherwise.

\revise{
The API Usage Plausibility (\aup) measures how effectively LLMs predict ``accurate'' APIs without considering their deprecation status. This metric is crucial because merely counting deprecated API usages (\dur) does not fully capture how LLMs manage API deprecations. For instance, a low number of deprecated API usages might be due to the prevalence of \others completions, suggesting influences beyond deprecation. To provide fair comparisons across models, prompts, and libraries, we calculate the Deprecated Usage Rate (\dur) based on AUP. A balanced relationship between AUP and DUR, characterized by a relatively high AUP and low DUR, indicates that LLMs effectively predict both up-to-date and ``accurate'' APIs. In contrast, an imbalance suggests either a lower rate of ``accurate'' predictions or a prevalence of outdated APIs. All the LLMs exhibited low AUPs due to the generation of many \others predictions, likely because many APIs (\eg PyTorch APIs) share similar usage contexts and offer flexible combinations. In addition, although these \others may provide alternative solutions for similar functionalities, they fall outside our focus on LLMs' handling of API deprecations. To clarify, we formally defined both \others and \aup to avoid confusion with terms like ``wrong'' or ``incorrect''. Nevertheless, we recognize the value of exploring this issue and leave a broader question for future work: how to effectively evaluate the functional correctness of generated completions, given the numerous alternative implementations for the same goal.
}
\section{Study Results}
We present the experimental results and key findings on the status quo and root causes of deprecated and replacement API usage in LLM-based code completion. As illustrated in Figure~\ref{fig:setup-overview}, the results are categorized into three detailed aspects: Model Perspective (RQ1), Prompt Perspective (RQ2), and Library Perspective (RQ3). For each RQ, the results and findings are presented by first showing the status quo through \textit{API Usage Plausibility} and \textit{Deprecated Usage Rate}, followed by providing an in-depth \textit{Root Cause Analysis}.

% Since the use of deprecated APIs can introduce unreliable factors such as software vulnerabilities~\cite{sigsoft/Wang0LC20}, it is essential to conduct an in-depth analysis to identify the root causes of deprecated API usage in LLM-based code completion. 

\subsection{RQ1: Model Perspective for Status Quo and Root Causes}\label{sec:rq1}
\subsubsection{Status Quo Analysis}
\begin{figure}
    \includegraphics[width=\columnwidth]{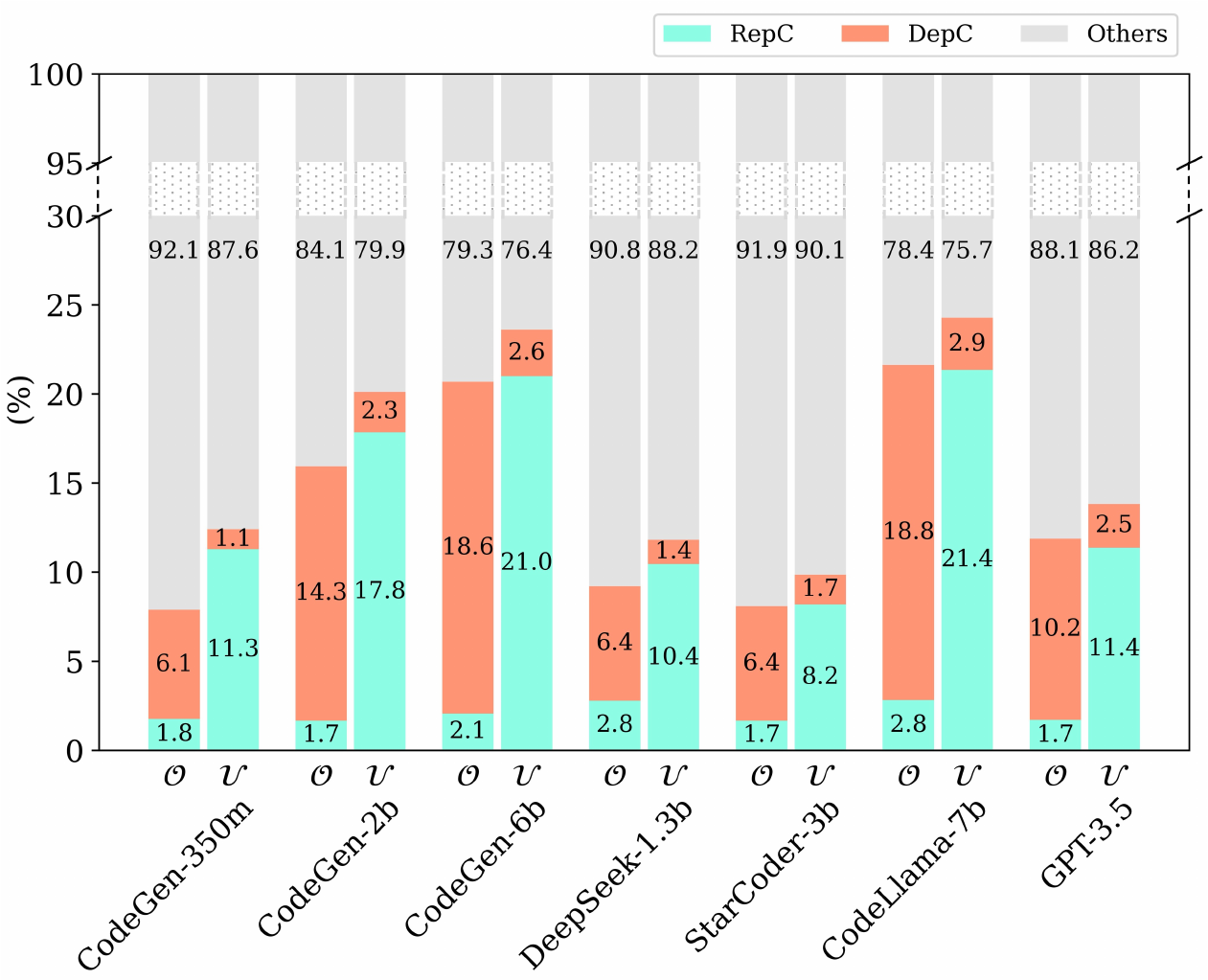}
    \vspace{-15pt}
    \caption{Distribution of \good, \bad, and \others Completions by LLMs for Prompts from $\mathcal{O}$ and $\mathcal{U}$}
    \label{fig:rq1-overall}
    \vspace{6pt}
\end{figure}

Figure~\ref{fig:rq1-overall} shows the distribution of \good and \bad completions by the different LLMs, and Table~\ref{tab:rq1-overall} presents the \aup and \dur metrics.

\textbf{API Usage Plausibility.}
The completion distribution and the low \aup highlight that all the LLMs faced challenges in predicting \plausible API usages for the given prompts.
The \aup of the LLMs for overall dataset (\ie All $=\mathcal{O}\cup\mathcal{U}$) ranges from \revise{9\% to 23\%}, indicating that a majority of predictions were \others. Among the LLMs, CodeLlama-7b achieved the highest \aup (23.4\%), while StarCoder2-3b had the lowest overall \aup (9.3\%). This may be attributed to the fact that StarCoder2 was not specifically fine-tuned on additional Python corpora, unlike other LLMs such as CodeLlama and CodeGen. 
Comparing the three versions of CodeGen suggests that the capacity of LLMs to predict \plausible API usages increased with model size (\ie 11.0\%, 28.8\%, and 23.6\% for CodeGen-350m, -2b, and -6b, respectively), given the model architecture and training data remain consistent. However, it's noteworthy that the largest LLM, GPT-3.5, did not achieve a high \aup (13.2\%), possibly due to the instruction used not being finely tuned with advanced prompt engineering techniques such as chain-of-thought (COT)~\cite{COT} and in-context learning (ICL)~\cite{ICL}.

\finding{
All the evaluated LLMs faced challenges in predicting \plausible API usages, with \aup ranging from 10\% to 30\%. Effectiveness of LLM-based code completion generally improved with model size and language-specific fine-tuning.
}

\begin{table}
    \caption{\aup and \dur for $\mathcal{O}$, $\mathcal{U}$, and  $\mathcal{O} \cup \mathcal{U}$}
    \vspace{-4pt}
    \label{tab:rq1-overall}
    \centering
    \renewcommand{\arraystretch}{1.2}
    \setlength{\tabcolsep}{4pt}
    \begin{tabularx}{0.95\columnwidth}{l|cYYYcYYY}
    \Xhline{2\arrayrulewidth}
    \multirow{2}{*}{\textbf{Model}} & & \multicolumn{3}{c}{\textbf{\aup (\%)}} & & \multicolumn{3}{c}{\textbf{\dur (\%)}} \\
    \cline{3-5}\cline{7-9}
    & & $\mathcal{O}$ & $\mathcal{U}$ & All & & $\mathcal{O}$ & $\mathcal{U}$ & All \\
    \Xhline{1.5\arrayrulewidth}
    CodeGen-350m & & \heatmapcell{7.9} & \heatmapcell{12.4} & \heatmapcell{11.0} & & \heatmapcell{77.7} & \heatmapcell{9.0} & \heatmapcell{24.9} \\
    CodeGen-2b & & \heatmapcell{15.9} & \heatmapcell{20.1} & \heatmapcell{18.8} & & \heatmapcell{89.6} & \heatmapcell{11.3} & \heatmapcell{32.6} \\
    CodeGen-6b & & \heatmapcell{20.7} & \heatmapcell{23.6} & \heatmapcell{22.7} & & \heatmapcell{90.0} & \heatmapcell{11.1} & \heatmapcell{34.2} \\
    DeepSeek-1.3b & & \heatmapcell{9.2} & \heatmapcell{11.8} & \heatmapcell{11.0} & & \heatmapcell{69.7} & \heatmapcell{11.6} & \heatmapcell{27.2} \\
    StarCoder-3b & & \heatmapcell{8.1} & \heatmapcell{9.9} & \heatmapcell{9.3} & & \heatmapcell{79.4} & \heatmapcell{17.0} & \heatmapcell{34.4} \\
    CodeLlama-7b & & \heatmapcell{21.6} & \heatmapcell{24.3} & \heatmapcell{23.4} & & \heatmapcell{86.9} & \heatmapcell{12.1} & \heatmapcell{34.2} \\
    GPT-3.5 & & \heatmapcell{11.9} & \heatmapcell{13.8} & \heatmapcell{13.2} & & \heatmapcell{85.7} & \heatmapcell{17.8} & \heatmapcell{37.4} \\
    \Xhline{2\arrayrulewidth}
    \end{tabularx}
\end{table}

\textbf{Deprecated Usage Rate.}
The distribution shown in Figure~\ref{fig:rq1-overall}, along with the \dur metric presented in Table~\ref{tab:rq1-overall}, indicates that all LLMs faced issues with using deprecated API usages.

The \dur of the LLMs for the overall dataset (\ie All $=\mathcal{O}\cup\mathcal{U}$) ranges from 25\% to 38\%, with larger models (\eg CodeGen-6b, CodeLlama-7b, and GPT-3.5) generally predicting more usages of deprecated APIs. %This trend may be attributed to the increased complexity and broader training data of larger models, which could include more deprecated API examples.
Considering the differences among the LLMs, CodeLlama-7b and CodeGen-6b demonstrated the best balance between \aup and \dur. They achieved significant improvements in \aup compared to other LLMs, with a comparable \dur of 34.2\%. Conversely, StarCoder2-3b and GPT-3.5 exhibited higher \dur (\ie 34.4\% and 37.4\%) despite having much lower \aup. These results indicate that the preference of LLMs for using deprecated or replacing APIs is not closely related to their capacity for predicting \plausible completions. 

\finding{
All the evaluated LLMs faced issues with deprecated API usages, with \dur ranging from 25\% to 38\%, and larger models exhibiting higher \dur. Among the LLMs, CodeLlama-7b and CodeGen-6b demonstrated the best balance between \aup and \dur.
}

\subsubsection{\revise{Potential Cause Discussion}}\label{sec:rq1:cause}
From the model perspective, the causes of deprecated API usages consist of two main points:

\noindent\textbf{Model Training:} \revise{As libraries evolve, both deprecated APIs and their replacements are commonly found in open-source code repositories. Since the training data for LLMs are primarily sourced from these repositories without filtering for deprecated APIs, they often include instances of deprecated API usage. While the training datasets for specific LLMs like CodeLlama and GPT-3.5 are not publicly available, preventing direct inspection, the distribution of deprecated API usages in open-source code repositories can serve as a proxy for understanding the nature of their training data. In this study, we identified 9,022 instances of deprecated API usage and 19,103 instances of their replacements in open-source code repositories. These coarse statistics suggest that a significant portion of the training data likely includes deprecated APIs. Moreover, prior studies have demonstrated a relationship between training data distribution and model output behavior~\cite{xie2023data,nijkamp2023codegen2}. When trained on data containing deprecated APIs, LLMs may ``memorize'' these APIs and associated usage patterns as part of their learned knowledge~\cite{LLM-KB,LLM-KB-Revisiting}. The different training datasets also led to different \aup and \dur of the LLMs.}

\noindent\textbf{Model Inference:} LLMs generated completions by predicting token probabilities based on their learned \emph{prior} knowledge (\eg memorized API usage contexts) and applying token selection strategies (\eg greedy search or beam search). Given certain contexts, LLMs were likely to predict deprecated API usages due to high token probabilities, without considering any \emph{posterior} API deprecation knowledge. 
\revise{To support this hypothesis, we calculated the generative likelihoods of both $dep$ and $rep$ for each prompt $pmpt$ across open-source LLMs by analyzing predicted token probabilities during decoding. To begin, we input $pmpt$ into an LLM, followed by sequentially feeding each token in $dep$ (or $rep$) into the model. Throughout this process, we recorded the predicted token probability distributions prior to each token in $dep$ (or $rep$) being input. Next, we extracted the predicted probabilities for tokens in $dep$ (or $rep$) from these distributions using their vocabulary indices. The generative likelihoods of $dep$ and $rep$ were then computed by summing the log probabilities of their tokens, a common method in language modeling~\cite{bengio2000neural}. We performed a paired t-test on these likelihoods. 
The t-statistic values indicate that for all six open-source LLMs, the likelihoods of deprecated APIs are significantly higher than those of replacing APIs for prompts from $\mathcal{O}$, whereas the opposite is true for prompts from $\mathcal{U}$. The $p$ values are approximately zero.}

\finding{
There are two primary reasons why LLMs predict deprecated APIs: the presence of deprecated API~usages in corpora during model training, and the absence of posterior knowledge about API deprecations during model inference.
}

\subsection{RQ2: Prompt Perspective for Status Quo and Root Causes}
\subsubsection{Status Quo Analysis}
Table~\ref{tab:rq1-overall} presents the \aup and \dur for prompts from the two datasets, \ie $\mathcal{O}$~and~$\mathcal{U}$.

\textbf{API Usage Plausibility.}
Between the prompts from the two different datasets, $\mathcal{O}$ and $\mathcal{U}$, there are some differences in \aup for all LLMs, with relative differences (\ie $(\aup^{\mathcal{U}} - \aup^{\mathcal{O}}) / \aup^{\mathcal{O}}$) ranging from \revise{12.5\% to 60.0\%}. This disparity may be attributed to the imbalance in the number of \buggy and \fixed functions in the LLMs' training corpora~\cite{Imbalance1,Imbalance2}. Indirect evidence for this is that, in this study, the \fixed functions collected from open-source code repositories were about twice as many as the \buggy~functions (\ie 19,103 \emph{vs.} 9,022), even though there was no collection preference. Given that LLMs were often trained on open-source code repositories, they likely learned more \fixed functions than \buggy functions, leading to better \aup for the $\mathcal{U}$ dataset. Additionally, larger LLMs showcased smaller \aup differences (\eg 16\% for CodeLlama-7b), possibly due to the better generalizability.

\finding{
The LLMs showcased difference in \aup between the two datasets $\mathcal{O}$ and $\mathcal{U}$. This difference is possibly attributed to the different distribution of \buggy and \fixed functions in the training corpora of LLMs.
}

\textbf{Deprecated Usage Rate.}
When considering $\mathcal{O}$ and $\mathcal{U}$ separately, all LLMs consistently demonstrated extremely high deprecated usage rates for $\mathcal{O}$ (70\%-90\% \dur) and relatively low rates for $\mathcal{U}$ (9\%-18\% \dur). This significant difference is also evident in the distribution of \good and \bad completions shown in Figure~\ref{fig:rq1-overall}. \revise{This observed discrepancy can be attributed to the tendency of LLMs to predict \textit{reference} APIs (i.e., the APIs used in the original functions) for most prompts in both the $\mathcal{O}$ and $\mathcal{U}$ datasets, reflecting the unique contextual characteristics of each prompt. To quantify this, we define an auxiliary metric, the \textit{Reference Usage Rate (RUR)}, as follows:
$$ \text{RUR} = \frac{\sum_{p \in \mathcal{P}} \mathbb{I}(\textbf{LLM}(p) = reference)}{\sum_{p \in \mathcal{P}} \mathbb{I}(\text{anno}(\textbf{LLM}(p)) \in \{ \text{\bad}, \text{\good}\})},$$
which measures the proportion of \textit{reference} APIs in plausible completions. RUR is equal to \dur for $\mathcal{O}$ and $(1 - \dur)$ for $\mathcal{U}$. Thus, no significant difference exists between $\mathcal{O}$ and $\mathcal{U}$ in terms of RUR, which is consistently within the 70\%-90\% range for both.} 
% In fact, for \plausible completions, the rate of \emph{reference} API usages (\ie the usages used in the original functions) equals \aup for $\mathcal{O}$, while it is $(1 - \aup)$ for $\mathcal{U}$. 
% Considering this rate, there is no noticeable difference between $\mathcal{O}$ and $\mathcal{U}$ (\ie around 70\%-90\% for both). This suggests that LLMs predicted the reference APIs for most prompts from both datasets, influenced by their differing characteristics. 
Nonetheless, the 9\%-18\% \dur for $\mathcal{U}$ indicates that LLMs still predicted the usage of deprecated APIs, even for the prompts from \fixed functions.

\finding{
The LLMs consistently exhibited a significant difference in \dur between the two datasets, with extremely high deprecated API usage rates for $\mathcal{O}$ (70\%-90\% \dur) and relatively low rates for $\mathcal{U}$ (9\%-18\% \dur).
}

\subsubsection{\revise{Potential Cause Discussion}}\label{sec:rq2:cause}
The contextual characteristics of the input completion prompts can significantly influence LLMs' use of deprecated APIs. Since the completions were generated based on the input prompts, specific contexts can lead LLMs to use deprecated APIs. As presented above, the LLMs showed significantly different \dur for prompts from $\mathcal{O}$ and $\mathcal{U}$. Upon comparing the prompts from the two datasets, we found that contextual characteristics of the prompts, such as specific variables and function calls, lead to the discrepancy of LLMs' predictions, \ie whether to use deprecated APIs or their replacements. 

\revise{
To quantitatively assess differences in contextual characteristics, we conducted lightweight contextual feature extraction for prompts from $\mathcal{O}$ and $\mathcal{U}$ using static analysis. Specifically, we grouped prompts according to their \textit{reference} API, denoted as $api$, with each group represented as $\mathcal{P}(api)$. For each prompt in $\mathcal{P}(api)$, we identified the function used to construct it (\cf Section~\ref{sec:matching}) and extracted the variable and function call names in this function. After analyzing all prompts within $\mathcal{P}(api)$, the extracted variable and function call names were compiled into a contextual feature set, represented by $\mathbf{Feat}(\mathcal{P}(api))$, for the prompt group. Through this approach, for each API mapping $dep \to rep$, we obtained two paired prompt groups $\mathcal{P}(dep) \subset \mathcal{O}$ and $\mathcal{P}(rep) \subset \mathcal{U}$, along with their feature sets, $\mathbf{Feat}(\mathcal{P}(dep))$ and $\mathbf{Feat}(\mathcal{P}(rep))$. We then assessed the contextual similarity between $\mathcal{P}(dep)$ and $\mathcal{P}(rep)$ using $\frac{|\mathbf{Feat}(\mathcal{P}(dep)) \cap \mathbf{Feat}(\mathcal{P}(rep))|}{|\mathbf{Feat}(\mathcal{P}(dep)) \cup \mathbf{Feat}(\mathcal{P}(rep))|}$.
The overall contextual similarity between $\mathcal{O}$ and $\mathcal{U}$ was determined by averaging the \text{cxt-sim} scores of paired prompt groups across all API mappings. The final similarity is 0.266. This low similarity highlights a significant difference in the contextual characteristics between the prompts from $\mathcal{O}$ and $\mathcal{U}$, influencing the model's reference decoding direction (\cf Section~\ref{sec:rq1:cause}).}

\finding{
The contextual characteristics of the input prompts, such as the defined variables and function calls, contribute a significant influence to the deprecated API usage.
}

\subsection{RQ3: Library Perspective for Status Quo and Root Causes}
We conducted a detailed analysis to examine the LLMs' completions across different libraries.

\subsubsection{Status Quo Analysis}
We observe how the \aup and \dur metrics vary with different libraries. 

\begin{figure}
    \includegraphics[width=1\columnwidth]{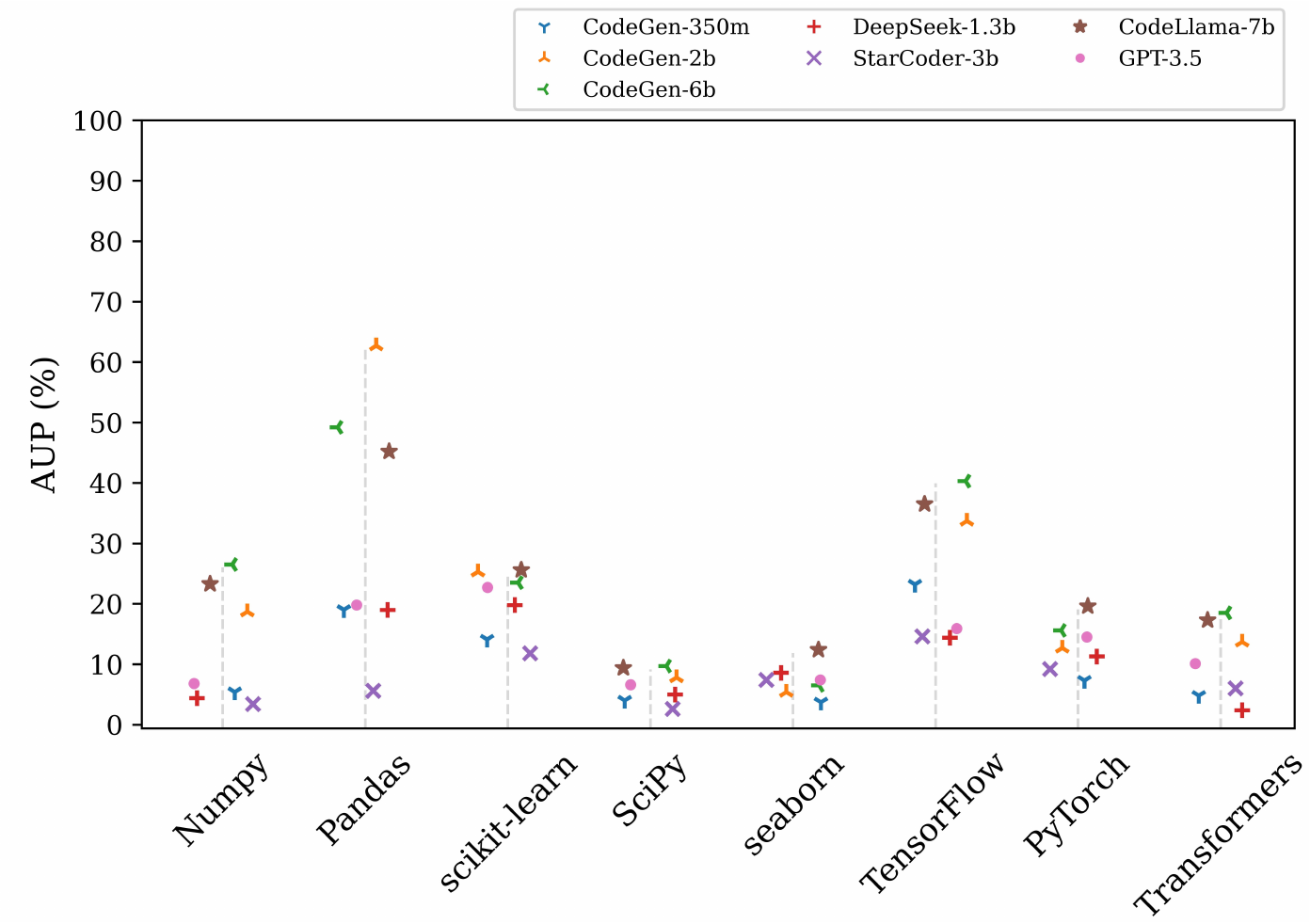}
    \vspace{-18pt}
    \caption{\aup by Different LLMs across Eight Libraries}
    \label{fig:rq1-library-aup}
\end{figure}

\textbf{API Usage Plausibility.}
The results of API usage plausibility are illustrated in the scatter plots depicted in Figure~\ref{fig:rq1-library-aup}, where each data point signifies the \aup of an LLM for a specific library.
Across the 8 libraries, most LLMs exhibited relatively low API usage plausibility for both $\mathcal{O}$ and $\mathcal{U}$, with \aup below 30\%. Notable exceptions were Pandas and TensorFlow, where CodeLlama-7b, CodeGen-6b, and CodeGen-2b (represented by symbols ``\mystar'',  ``\trialeft'', and ``\triaup'', respectively) achieved better \aup (around or greater than 40\%). This aligns with the results presented in Model Perspective, where CodeLlama-7b, CodeGen-6b, and CodeGen-2b achieved the best results for the overall dataset. On the other hand, among the libraries, completion prompts from SciPy and seaborn posed the most difficulty for LLMs in predicting \plausible API usages, with \aup consistently below 15\%. 

\finding{
Most LLMs exhibited relatively low API usage plausibility across the 8 libraries, with \aup below 30\%. CodeLlama-7b, CodeGen-6b, and CodeGen-2b achieved better \aup for Pandas (about 45\%-65\%) and TensorFlow (around 40\%).
}

\begin{figure}
    % \vspace{-8pt}
    \includegraphics[width=1\columnwidth]{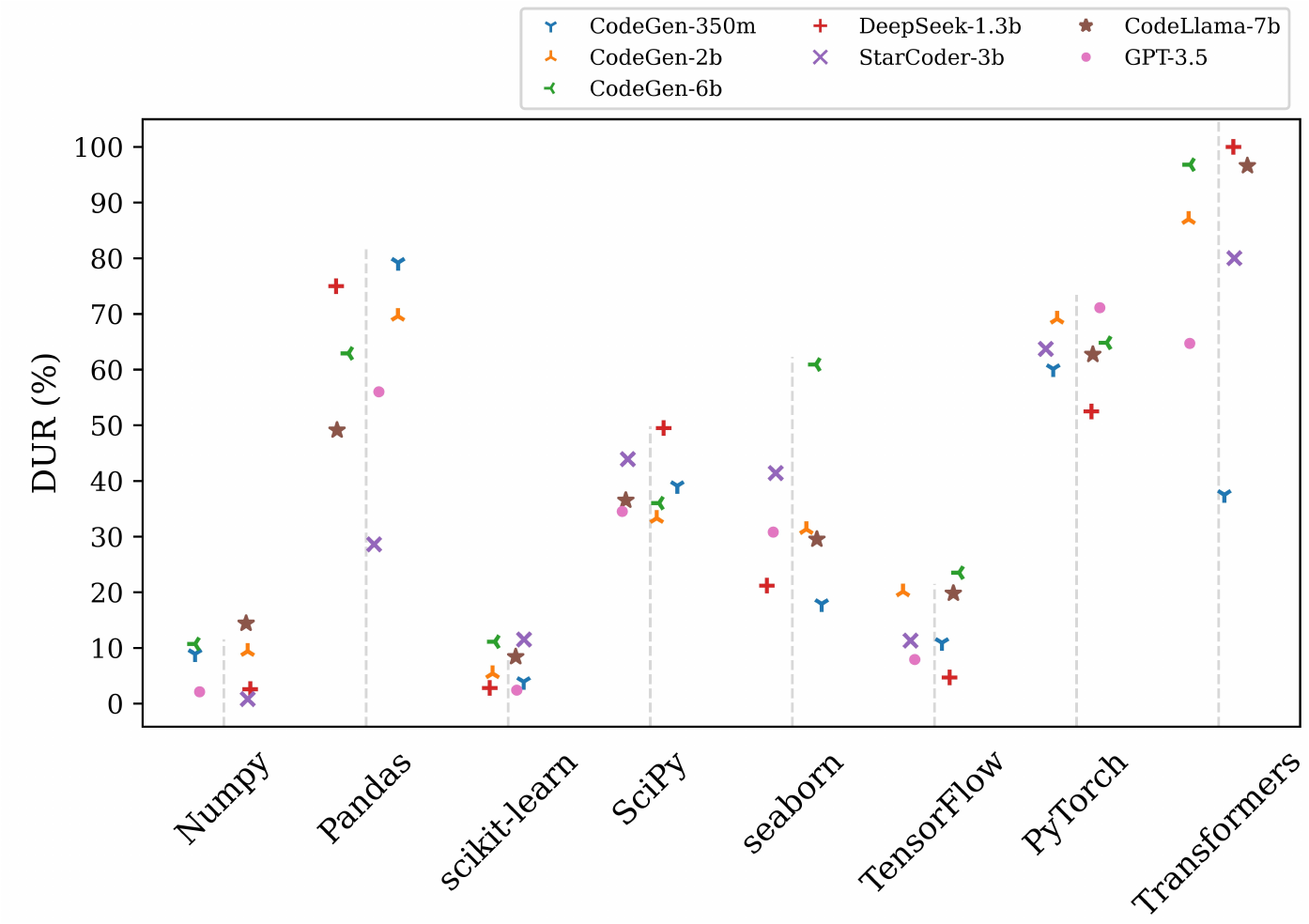}
    \vspace{-18pt}
    \caption{\dur by Different LLMs across Eight Libraries}
    \label{fig:rq1-library-dur}
\end{figure}

\textbf{Deprecated Usage Rate.}
The results are presented in the scatter plots shown in Figure~\ref{fig:rq1-library-dur}, where each data point represents the \dur of a particular LLM for a specific library. The results reveal significant differences in the usage of deprecated APIs across the libraries, with \dur ranging from approximately 0\% to 100\%. Specifically, LLMs generally showed low \dur (around or below 20\%) for Numpy, scikit-learn, and TensorFlow. In contrast, LLMs exhibited consistently high \dur for SciPy (approximately 30\%-50\%) and PyTorch (approximately 50\%-70\%), and unstable \dur for Pandas (approximately 30\%-80\%), seaborn (approximately 20\%-60\%), and Transformers (approximately 40\%-100\%). 

\finding{
LLMs showed significant differences in the usage of deprecated APIs across various libraries, with \dur ranging from approximately 0\% to 100\%. LLMs exhibited consistently high \dur for SciPy and PyTorch, and unstable \dur for Pandas and Transformers. 
}

\subsubsection{\revise{Potential Cause Discussion}}
After analyzing the completion prompts and LLMs' predictions, we found that the \aup differences between these libraries were primarily due to the characteristics of the API usage context. 
More specifically, the usage contexts of certain APIs followed common patterns and were surrounded by related APIs, allowing advanced LLMs like CodeLlama-7b to infer the desired API functionality based on the completion prompt. For example, many utility APIs in TensorFlow, such as the deprecated \inlinecode{tensorflow.compat.v1.initialize\_all\_variables} and its replacement \inlinecode{tensorflow.compat.v1.global\_variables\_initializer}, were often used in recognizable patterns that were easier for LLMs to predict. Conversely, some APIs were used more flexibly in diverse contexts and lacked obvious combinations with other APIs, leading to low \aup for LLMs in predicting such APIs. For instance, many APIs in SciPy, like the deprecated \inlinecode{scipy.misc.comb} and its replacement \inlinecode{scipy.special.comb}, can be used in diverse contexts to produce combinations for a data sequence. Since the data sequence can originate from numerous sources and be structured in various ways (\eg Numpy array and Pandas Series), it is challenging for LLMs to predict these APIs based on the completion prompt without additional hints.

The characteristics of API deprecations during library evolution also significantly impacted the use of deprecated APIs in LLM-based code completions. Some APIs were deprecated due to simple package refactoring, leading to similar usage patterns for the deprecated APIs and their replacements. For example, in the version 1.9.0 release of PyTorch, many APIs for tensor linear algebra were moved from the \inlinecode{torch} package to the \inlinecode{torch.linalg} package without changes to the API parameters or usage patterns (\eg \inlinecode{torch.lstsq} $\to$ \inlinecode{torch.linalg.lstsq}). In such cases, LLMs found it more difficult to distinguish between deprecated APIs and their replacements, resulting in more frequent use of deprecated APIs in the predicted completions.
\revise{To further validate this attribution, we gathered prompts associated with each library from $\mathcal{O}$ and $\mathcal{U}$, denoted as $\mathcal{O}^{l}$ and $\mathcal{U}^{l}$. We then performed feature extraction and calculated the contextual similarity between $\mathcal{O}^{l}$ and $\mathcal{U}^{l}$ following the same approach as in Section~\ref{sec:rq2:cause}. A higher contextual similarity indicates that $\mathcal{O}^{l}$ and $\mathcal{U}^{l}$, before and after API deprecation, share more common contextual characteristics and usage patterns. Higher $\mathcal{O}^{l}$-$\mathcal{U}^{l}$ contextual similarity suggests that it is more challenging for LLMs to discern whether to use deprecated or replacing APIs based on the given prompts. Table~\ref{tab:rq3-ctx-sims} shows the relationship between $\mathcal{O}^{l}$-$\mathcal{U}^{l}$ contextual similarity and \dur ranges across the eight libraries, indicating that the \dur range generally increases with higher $\mathcal{O}^{l}$-$\mathcal{U}^{l}$ contextual similarity. }

\begin{table}
    \caption{\revise{Relationship between $\mathcal{O}^{l}$-$\mathcal{U}^{l}$ Contextual Similarity and \dur Range across Eight Libraries}}
    \label{tab:rq3-ctx-sims}
    \centering
    \renewcommand{\arraystretch}{1.2}
    \setlength{\tabcolsep}{12pt}
    \begin{tabularx}{0.75\columnwidth}{l|rr}
    \Xhline{2\arrayrulewidth}
    \textbf{Library} & \textbf{Similarity} & \textbf{\dur Range} \\
    \Xhline{1.5\arrayrulewidth} 
    Numpy & 0.056 & $<$ 20\% \\
    Pandas & 0.193 & 30\%-80\% \\
    scikit-learn & 0.060 & $<$ 20\% \\
    SciPy & 0.094 & 30\%-50\% \\
    seaborn & 0.054 & 20\%-60\% \\
    TensorFlow & 0.078 & $<$ 25\% \\
    PyTorch & 0.126 & 50\%-70\% \\
    Transformers & 0.672 & 40\%-100\% \\
    \Xhline{2\arrayrulewidth}
    \end{tabularx}
\end{table}

\finding{
The characteristics of API deprecations during library evolution significantly impacted the use of deprecated APIs in LLM-based code completions. Minor changes between deprecated APIs and their replacements, such as simple package refactoring, often led to more pronounced issues with deprecated API usages.
}
\section{Lightweight Mitigations}
Based on the findings regarding the causes of deprecated API usage, we proposed two lightweight mitigation approaches, which can serve as baselines for future investigation. 

\subsection{Motivation}
As analyzed in the RQs, the causes of deprecated API usage in LLM-based completions can be attributed to Model Training, Model Inference, Prompt, and Library aspects. In this section, we further explore the feasibility of mitigating deprecated API usage issues from these four perspectives.

During model training, a direct mitigation is to clean~up the code containing deprecated APIs from the training corpora. However, this is impractical due to the constant library~evolution. New API deprecations would cause repeatedly training corpora rebuilding and model retraining. From the library perspective, we cannot control the evolution~of~the library and API deprecation. Therefore,~mitigation~should focus on decoding strategies in model inference and prompt~engineering.

\subsection{Two Lightweight Approaches}
Our basic idea is illustrated in Algorithm~\ref{alg:fixing}. Given an LLM, its generation process can generally be formulated as follows:
$$\mathbb{T}^O = \proc{LLM}(\mathbb{T}^I),$$
where $\mathbb{T}^I$ and $\mathbb{T}^O$ are the input token sequence and output token sequence, respectively. In the context of LLM-based code completion, $\mathbb{T}^I$ corresponds to the input prompt $pmpt$, and $\mathbb{T}^O$ corresponds to the predicted completion $comp$ (line 1). When $comp$ contains a deprecated API $dep$ (line 3), we need to perform fixing approaches, \ie the \proc{Fix} procedure, to replace $dep$ with the corresponding replacement $rep$ (line 4). Note that during the \proc{Contains} procedure in line 3, alias resolution is conducted similarly to the process in~Section~\ref{sec:setup:ccpc:matching}.

\begin{algorithm}[tp]
\caption{Deprecation-Aware Code Completion}
\label{alg:fixing}
\DontPrintSemicolon
    \KwInput{Prompt $pmpt$, API Mappings $\mathcal{M}$}
    \KwOutput{Completion $comp$}
    $comp \gets \proc{LLM}(pmpt)$ \\
    \For{$(dep \to rep) \in \mathcal{M}$}
    {
        \If{$\emph{\proc{Contains}}(comp, dep)$}
        {
            $comp \gets \proc{Fix}(pmpt, comp, dep, rep)$  \\  
            \textbf{break}
        }
    }
\end{algorithm}

The \proc{Fix} procedure includes reconstructing input, through either (i) replacing the deprecated API tokens or (ii) inserting additional replacing prompts, and then regenerating~output:

\begin{itemize}[leftmargin=*]
    \item \textbf{Approach 1 - \textsc{ReplaceAPI}: Replacing Deprecated API Tokens then Regenerating.}
    As shown in Algorithm~\ref{alg:replace}, \proc{ReplaceDep} removes the tokens corresponding to $dep$ and any subsequent tokens from $comp$, and then appends the tokens of $rep$ (line 2). This results in a prefix $prefix$ that includes $rep$. The $prefix$ is concatenated with the $pmpt$, and the LLM generates the suffix (line 3), \ie arguments for $rep$ and the remaining tokens to complete the code line. This concatenation forms the fixed completion $comp^*$.

    \item \textbf{Approach 2 - \textsc{InsertPrompt}: Inserting Additional Replacing Prompt then Regenerating.}
    As shown in Algorithm~\ref{alg:insert}, \proc{CreateDepPmpt} constructs an additional replacing prompt $pmpt'$ (line 2), formatted as inline comments to guide the LLM to use $rep$ instead of $dep$. By expressing the replacing instruction into inline comments, the re-written prompt (\ie $pmpt \oplus pmpt'$) can be naturally processed by the LLM to continue writing the code. The replacing prompt is structured as follows, where ``\texttt{\small\{comp\}}'', ``\texttt{\small\{dep\}}'', and ``\texttt{\small\{rep\}}'' are placeholders for the original completion, deprecated API, and replacing API, respectively, and ``\texttt{\small...}'' represents the indentation to ensure syntax~correctness. 
\begin{lstlisting}[language=python,breaklines=true]
...# {comp}
...# {dep} is deprecated, use {rep} instead and revise the return value and arguments.
\end{lstlisting}
    The created $pmpt'$ is then concatenated with $pmpt$ and fed into the LLM to generate a new completion $comp^*$  (line 3). 
\end{itemize}

\begin{algorithm}[tp]
\caption{Approach 1 - \textsc{ReplaceAPI}}
\label{alg:replace}
\DontPrintSemicolon
    \SetKwProg{Fn}{Procedure}{:}{}
    \Fn{\emph{\proc{Fix}}{($pmpt, comp, dep, rep$)}} {
        $prefix \gets \proc{ReplaceDep}(comp, dep, rep)$ \\
        $suffix \gets \proc{LLM}(pmpt \oplus prefix)$ \\
        $comp^* \gets prefix \oplus suffix$\\
        \textbf{return} $comp^*$
    }
\end{algorithm}

\begin{algorithm}
\caption{Approach 2 - \textsc{InsertPrompt}}
\label{alg:insert}
\DontPrintSemicolon
    \SetKwProg{Fn}{Procedure}{:}{}
    \Fn{\emph{\proc{Fix}}{($pmpt, comp, dep, rep$)}} {
        $pmpt' \gets \proc{CreateRepPmpt}(comp, dep, rep)$ \\
        $comp^* \gets\proc{LLM}(pmpt \oplus pmpt')$ \\
        \textbf{return} $comp^*$
    }
\end{algorithm}

\subsection{Evaluation}
We conducted experiments to address the following research question: 
\begin{itemize}[leftmargin=12pt]
    \item \textbf{RQ4:} How effectively can the proposed approaches fix deprecated API usage in completions? %How accurate are the completions generated by the proposed approaches compared to ground-truth completions?
\end{itemize}

\subsubsection{Evaluation Procedure}
For each LLM, we selected up-to-dated samples from $\mathcal{U}$ where the LLM predicted \emph{\bad} completions using deprecated APIs, \ie
$$\mathcal{T} = \{(pmpt, dep \to rep) \in \mathcal{U} : \text{anno}(\proc{LLM}(pmpt)) = \bad\},$$
to constitute the evaluation data. These samples were chosen because they have corresponding ground-truth completions (e.g., the line following the prompt in Figure~\ref{fig:prompt}), which are essential for assessing the effectiveness of the proposed fixing approaches. For each sample, we employed the deprecation-aware code completion illustrated in Algorithm~\ref{alg:fixing} with the two fixing approaches \textsc{ReplaceAPI} and \textsc{InsertPrompt} to prompt the LLM to generate a completion. We used the following three metrics to assess their effectiveness:
\begin{itemize}[leftmargin=15pt]
    \item \textbf{Fixed Rate (FR)}: This metric indicates the proportion of \good completions predicted by the fixing approaches.
    \item \textbf{Edit Similarity (ES)}~\cite{EditSim}: This metric measures the similarity between the predicted completions and the ground-truth completions by analyzing the edit operations required to transform one into the other. 
    \item \textbf{Exact Match (EM)}: This metric calculates the rate of predicted completions that exactly match the ground-truth completions after normalizing the return values of function calls (\ie replacing each element in return value with ``\_'').
\end{itemize}

\subsubsection{Results and Analysis}
The evaluation results of the proposed fixing approaches are presented in Table~\ref{tab:fixing}. 

\begin{table}[]
    \footnotesize
    \caption{Evaluation Results of Proposed Approaches}
    \vspace{-3pt}
    \label{tab:fixing}
    \centering
    \renewcommand{\arraystretch}{1.2}
    \setlength{\tabcolsep}{2pt}
    \begin{tabular}{l|cccccccc}
        \Xhline{2\arrayrulewidth}
        \multirow{2}{*}{\textbf{Model}} & & \multicolumn{3}{c}{\textbf{\textsc{ReplaceAPI}}}   & & \multicolumn{3}{c}{\textbf{\textsc{InsertPrompt}}} \\
        \cline{3-5}\cline{7-9}
                                        & & \textbf{FR} \small{(\%)} & \textbf{ES} \small{(\%)} & \textbf{EM} \small{(\%)} & & \textbf{FR} \small{(\%)} & \textbf{ES} \small{(\%)} & \textbf{EM} \small{(\%)} \\
        \Xhline{1.5\arrayrulewidth}
        CodeGen-350m                    & & 92.1           & 82.3           & 30.8           & & 25.7           & 58.7           & 8.9                \\
        CodeGen-2b                      & & 88.2           & 84.6           & 38.8           & & 66.1           & 66.0           & 23.3               \\
        CodeGen-6b                      & & 85.2           & 85.3           & 43.5           & & 77.4           & 72.9           & 35.3               \\
        DeepSeek-1.3b                   & & 99.6           & 80.9           & 31.7           & & 93.5           & 77.7           & 24.4               \\
        StarCoder-3b                    & & 90.9           & 85.0           & 42.2           & & 85.3           & 72.2           & 29.1               \\
        CodeLlama-7b                    & & 99.5           & 85.7           & 48.1           & & 95.5           & 82.0           & 43.3               \\
        GPT-3.5                         & & --             & --             & --             & & 97.2           & 76.2           & 20.5               \\
        \Xhline{2\arrayrulewidth}
    \end{tabular}
\end{table}

\textbf{\textsc{ReplaceAPI}.}
Using the \textsc{ReplaceAPI} fixing~approach (Algorithm~\ref{alg:replace}), all the LLMs achieve high fixed rates (FR), with values exceeding 85\%. Failures in fixing are mainly due to syntax errors or incorrect function calls caused by erroneous tokens following the replaced APIs. For example, consider a \emph{\bad} completion: ``meta\_graph\_def = tf.saved\_model.loader.load(...)'' predicted by the original completion procedure~(Algorithm~\ref{alg:fixing}, line 1). \textsc{ReplaceAPI} replaces the deprecated API \inlinecode{tf.saved\_model.loader.load} with its replacement \inlinecode{tf.saved\_model.load}, producing a $prefix$ of ``meta\_graph\_def=tf.saved\_model.load'' (line 2 of Algorithm~\ref{alg:replace}). However, CodeGen-2b then predicts a $suffix$ of ``\_meta\_graph\_def(...)'' (line 3 of Algorithm~\ref{alg:replace}), resulting in an erroneous function call: \inlinecode{tf.saved\_model.load\_meta\_graph\_def()}. This issue arises because the replacement operation in \textsc{ReplaceAPI} can disrupt the naturalness of the code context~\cite{Naturalness} and the LLMs' decoding process. An additional interesting finding is that for the three versions of CodeGen, the FR decreases as the model size increases,~suggesting that larger models might be more sensitive to interventions~during~decoding.

The completions generated by LLMs using the \textsc{ReplaceAPI} approach exhibit high ES of over 80\%, and EM rates between 30\% and 50\%, with these rates increasing alongside model size. The inaccuracies in the completions often involve incorrect return values and arguments for the replacing APIs. Incorrect return values arise because the replacing API might include different elements compared to the deprecated API, and the \textsc{ReplaceAPI} approach cannot resolve such inconsistencies in the $prefix$ (line 2 of Algorithm~\ref{alg:replace}). The incorrect arguments are primarily due to the LLMs' limitations in correctly utilizing replacing APIs, especially those with complex argument lists.

\textbf{\textsc{InsertPrompt}.}
When applying the \textsc{InsertPrompt} fixing approach (Algorithm~\ref{alg:insert}), the LLMs exhibited significantly varied fixed rates, ranging from 25.7\% to 97.2\%. This variation suggests that larger models generally possess a stronger capacity to interpret and utilize inserted prompts formatted as inline comments. An exception to this trend is DeepSeek-1.3b, which achieved a notably high FR of 93.5\%. This success can be attributed to using the \texttt{deepseek-coder-1.3b-instruct} version, which has robust zero-shot instruction-following capabilities. Moreover, the FR differences among various LLMs highlight their sensitivity to prompt construction~\cite{PromptSensitivity1,PromptSensitivity2}, which indicates that different LLMs may require specialized additional prompts in \textsc{InsertPrompt} for optimal performance.

Considering edit similarity and exact match, the completions generated by GPT-3.5 and DeepSeek-1.3b using the \textsc{InsertPrompt} approach do not perform as well as their fixed rates suggest. Despite being fine-tuned with an instruct-tuning corpus, they are not specifically fine-tuned on Python code. As a result, they can follow the instructions in the additional prompt to use replacing APIs but struggle to use those APIs correctly.

\textbf{Comparison.}
The comparison between \textsc{ReplaceAPI} and \textsc{InsertPrompt} suggests that direct interventions in the decoding process are more effective than zero-shot prompt engineering. However, the results also reveal the potential of \textsc{InsertPrompt}. 
First, \textsc{ReplaceAPI} cannot be applied to black-box LLMs like GPT-3.5, as their decoding processes cannot be controlled by users. Second, the additional prompt employed by \textsc{InsertPrompt} was not carefully tuned for each LLM and was performed in a zero-shot manner. In the future, fine-tuning the LLMs with instructions specifically designed for fixing deprecated usage could enhance effectiveness.

\finding{
\textsc{ReplaceAPI} effectively addresses deprecated API usage for all open-source LLMs, achieving fix rates exceeding 85\%. The fixed completions demonstrate acceptable accuracy. While \textsc{InsertPrompt} does not achieve sufficient effectiveness and accuracy in fixing completions containing deprecated API usage, it provides future potential of exploration.
}

\section{Discussion}

\subsection{Implications}
% We provide implications for future research on the synergy of library evolution and LLM-driven software development.

\textbf{Validating LLM-Generated Code Completions and Issuing Alerts for Deprecated API Usages.}
Our evaluation study reveals that LLMs frequently use deprecated APIs during code completion. Such deprecated API usages can be easily overlooked~\cite{sigsoft/Wang0LC20}, potentially introducing bugs or security vulnerabilities into software projects. Therefore, implementing a validation mechanism for deprecated API usage in LLM-generated code completions is crucial to ensure the reliability of the code. Such a validation mechanism can be further integrated into the post-processing of LLM-based code completion, such as issuing alerts to developers.

\textbf{Fixing and Updating Outdated API Knowledge in LLMs by Model-Level Repair.}
The current fixing approaches mitigate the issues by intervening in decoding process and rewriting prompts, without addressing the outdated knowledge about deprecated API usages stored in the LLMs. Given the constant evolution of libraries, lightweight model repair techniques are potential solutions for fixing outdated knowledge. \textit{Model Editing} is one such direction, which can be categorized into the following main categories: Memory-based approaches~\cite{Mitchell22Memory,Murty22Fixing,Madaan22Memory}, Locating-then-editing approaches~\cite{Meng22Locating,Meng23Editing,Li24PMET}, and Meta-learning approaches~\cite{Cao21Editing,Mitchell22Fast}. Compared to the proposed fixing approaches, model editing can directly update the outdated knowledge about deprecated API usages, even incorporating the information of entirely new replacing APIs (\ie the replacing APIs introduced after model training and unseen in the training corpora) into the LLMs.

\textbf{Leveraging Retrieval-Augmented Generation for Up-to-dated Code Completion.}
As discussed in our study findings, a key cause of the deprecated API usage in LLM-based completions is the lack of posterior knowledge about API deprecations. Retrieval-Augmented Generation (RAG) is a suitable technique that can perfectly align with the need for posterior knowledge~\cite{Lewis20RAG}. We can explore the possibility of adopting RAG to mitigate deprecated API usage in LLM-based code completion by retrieving related knowledge pieces, such as documentation and usage examples.

\textbf{Designing Agent \& Multi-Agent Systems for Incorporating Library Evolution into LLM-Driven Software Development.}
In modern software development driven by LLMs, the issues brought by library evolution are encountered not only in code completion. With advancements in AI agents~\cite{gronauer2022multi,talebirad2023multi,ellis2024smacv2}, we can potentially develop autonomous agents or multi-agent systems capable of automatically discovering deprecated API usages, identifying correct replacements, upgrading dependent libraries, and fixing the code~\cite{wang2024towards}. To ensure comprehensive recognition of deprecated API usage and up-to-date fixes, we should design an effective multi-agent collaboration pipeline. One agent should scan the generated code to identify all pieces related to API usage. Another agent should continuously fetch information online, checking the latest official API documentation to aid in discovery and correction. Finally, a dedicated agent should be responsible for implementing the necessary library upgrades and code fixes. 
% \revise{A key technical consideration is the integration of existing analysis tools, such as automated API mapping collection~\cite{Huang2021}, to fulfill the functional requirements effectively.} Such an agent system can fundamentally address the issues discussed in this article.

\subsection{Threats to Validity}
\textbf{Internal Threats.}
The primary threat to our study is the soundness of the static analysis used for function location and result annotation. Given that Python is a dynamic programming language, the lightweight object type resolution and alis resolution we employed may have missed some function calls of deprecated and replacing APIs during the matching process. \hkf{Nevertheless, they did not contribute bias to deprecated APIs or replacing APIs in our extracted function calls.} In the future, we aim to address this issue by implementing advanced type inference techniques. Additionally, our study currently focuses on function-level API deprecation, overlooking parameter-level deprecations. Future research would investigate a broader type of deprecated APIs to provide more comprehensive~analysis.

\textbf{External Threats.}
A primary threat lies in the choice of Python libraries and LLMs. To mitigate this threat, we reused libraries examined in previous studies and introduced three popular deep learning libraries to ensure diversity and timeliness. For the LLMs, we selected models covering various architectures, model sizes, training corpora, and training~strategies to ensure the generalizability of our findings. Another external threat is that the study was conducted solely on the Python language, which may limit the applicability of our findings to other languages such as Java and C\#. In the future, we plan to explore the impact of library evolution on LLM-based code completion across a broader range~of~programming~languages.
\section{Conclusion}
In this work, we conducted an evaluation study to investigate the statuses and causes of deprecated API usages in LLM-based code completion. The study results all evaluated LLMs encounter challenges in predicting \plausible API usages and face issues with deprecated API usages, influenced by the distinct code context characteristics of the prompts and the characteristics of API deprecations during the evolution of these libraries. We propose two lightweight fixing approaches to mitigate the deprecated API usages and can serve as baselines for future research. We also provide implications for the research directions for the combination of library evolution and LLM-driven code completion and software development. We released the code and data of our study at the website~\cite{replication}.

\section*{Acknowledgement}
This research project is supported by the National Key R\&D Program of China (2023YFB4503805), the National Research Foundation, Singapore, and the Cyber Security Agency under its National Cybersecurity R\&D Programme (NCRP25-P04-TAICeN), DSO National Laboratories under the AI Singapore Programme (AISG2-GC-2023-008), National Natural Science Foundation of China under Grant No. 62402342, Shanghai Sailing Program (No. 24YF2749500), and the Ministry of Education, Singapore, under its Academic Research Fund Tier 1 (RG96/23).

\clearpage
\bibliographystyle{IEEEtran}
\balance
\bibliography{main.bib}

\begin{thebibliography}{10}
\providecommand{\url}[1]{#1}
\csname url@samestyle\endcsname
\providecommand{\newblock}{\relax}
\providecommand{\bibinfo}[2]{#2}
\providecommand{\BIBentrySTDinterwordspacing}{\spaceskip=0pt\relax}
\providecommand{\BIBentryALTinterwordstretchfactor}{4}
\providecommand{\BIBentryALTinterwordspacing}{\spaceskip=\fontdimen2\font plus
\BIBentryALTinterwordstretchfactor\fontdimen3\font minus \fontdimen4\font\relax}
\providecommand{\BIBforeignlanguage}[2]{{%
\expandafter\ifx\csname l@#1\endcsname\relax
\typeout{** WARNING: IEEEtran.bst: No hyphenation pattern has been}%
\typeout{** loaded for the language `#1'. Using the pattern for}%
\typeout{** the default language instead.}%
\else
\language=\csname l@#1\endcsname
\fi
#2}}
\providecommand{\BIBdecl}{\relax}
\BIBdecl

\bibitem{Codex}
\BIBentryALTinterwordspacing
M.~Chen, J.~Tworek, H.~Jun, Q.~Yuan, H.~P. de~Oliveira~Pinto, J.~Kaplan, H.~Edwards, Y.~Burda, N.~Joseph, G.~Brockman, and Others, ``Evaluating large language models trained on code,'' \emph{CoRR}, vol. abs/2107.03374, 2021. [Online]. Available: \url{https://arxiv.org/abs/2107.03374}
\BIBentrySTDinterwordspacing

\bibitem{InCoder}
D.~Fried, A.~Aghajanyan, J.~Lin, S.~Wang, E.~Wallace, F.~Shi, R.~Zhong, W.-t. Yih, L.~Zettlemoyer, and M.~Lewis, ``Incoder: A generative model for code infilling and synthesis,'' \emph{arXiv preprint arXiv:2204.05999}, 2022.

\bibitem{CodeGen}
E.~Nijkamp, B.~Pang, H.~Hayashi, L.~Tu, H.~Wang, Y.~Zhou, S.~Savarese, and C.~Xiong, ``Codegen: An open large language model for code with multi-turn program synthesis,'' \emph{arXiv preprint arXiv:2203.13474}, 2022.

\bibitem{StarCoder}
R.~Li, L.~B. Allal, Y.~Zi, N.~Muennighoff, D.~Kocetkov, C.~Mou, M.~Marone, C.~Akiki, J.~Li, J.~Chim \emph{et~al.}, ``Starcoder: may the source be with you!'' \emph{arXiv preprint arXiv:2305.06161}, 2023.

\bibitem{CodeLlama}
B.~Rozi{\`{e}}re, J.~Gehring, F.~Gloeckle, S.~Sootla, I.~Gat, X.~E. Tan, Y.~Adi, J.~Liu, T.~Remez, J.~Rapin, A.~Kozhevnikov, I.~Evtimov, J.~Bitton, M.~Bhatt, C.~Canton{-}Ferrer, A.~Grattafiori, W.~Xiong, A.~D{\'{e}}fossez, J.~Copet, F.~Azhar, H.~Touvron, L.~Martin, N.~Usunier, T.~Scialom, and G.~Synnaeve, ``Code llama: Open foundation models for code,'' \emph{CoRR}, vol. abs/2308.12950, 2023.

\bibitem{DeepSeek-Coder}
D.~Guo, Q.~Zhu, D.~Yang, Z.~Xie, K.~Dong, W.~Zhang, G.~Chen, X.~Bi, Y.~Wu, Y.~K. Li, F.~Luo, Y.~Xiong, and W.~Liang, ``Deepseek-coder: When the large language model meets programming - the rise of code intelligence,'' \emph{CoRR}, vol. abs/2401.14196, 2024.

\bibitem{chen2025deep}
X.~Chen, X.~Hu, Y.~Huang, H.~Jiang, W.~Ji, Y.~Jiang, Y.~Jiang, B.~Liu, H.~Liu, X.~Li \emph{et~al.}, ``Deep learning-based software engineering: progress, challenges, and opportunities,'' \emph{Science China Information Sciences}, vol.~68, no.~1, pp. 1--88, 2025.

\bibitem{svyatkovskiy2020intellicode}
A.~Svyatkovskiy, S.~K. Deng, S.~Fu, and N.~Sundaresan, ``Intellicode compose: Code generation using transformer,'' in \emph{Proceedings of the 28th ACM joint meeting on European software engineering conference and symposium on the foundations of software engineering}, 2020, pp. 1433--1443.

\bibitem{le2022coderl}
H.~Le, Y.~Wang, A.~D. Gotmare, S.~Savarese, and S.~C.~H. Hoi, ``Coderl: Mastering code generation through pretrained models and deep reinforcement learning,'' \emph{Advances in Neural Information Processing Systems}, vol.~35, pp. 21\,314--21\,328, 2022.

\bibitem{wang2024teaching}
C.~Wang, J.~Zhang, Y.~Feng, T.~Li, W.~Sun, Y.~Liu, and X.~Peng, ``Teaching code llms to use autocompletion tools in repository-level code generation,'' \emph{arXiv preprint arXiv:2401.06391}, 2024.

\bibitem{zeng2022extensive}
Z.~Zeng, H.~Tan, H.~Zhang, J.~Li, Y.~Zhang, and L.~Zhang, ``An extensive study on pre-trained models for program understanding and generation,'' in \emph{Proceedings of the 31st ACM SIGSOFT international symposium on software testing and analysis}, 2022, pp. 39--51.

\bibitem{wang2024tiger}
C.~Wang, J.~Zhang, Y.~Lou, M.~Liu, W.~Sun, Y.~Liu, and X.~Peng, ``Tiger: A generating-then-ranking framework for practical python type inference,'' \emph{arXiv preprint arXiv:2407.02095}, 2024.

\bibitem{wang2023boosting}
C.~Wang, J.~Liu, X.~Peng, Y.~Liu, and Y.~Lou, ``Boosting static resource leak detection via llm-based resource-oriented intention inference,'' \emph{arXiv preprint arXiv:2311.04448}, 2023.

\bibitem{li2024llm}
J.~Li, Z.~Dong, C.~Wang, H.~You, C.~Zhang, Y.~Liu, and X.~Peng, ``Llm based input space partitioning testing for library apis,'' \emph{arXiv preprint arXiv:2501.05456}, 2024.

\bibitem{fan2023automated}
Z.~Fan, X.~Gao, M.~Mirchev, A.~Roychoudhury, and S.~H. Tan, ``Automated repair of programs from large language models,'' in \emph{2023 IEEE/ACM 45th International Conference on Software Engineering (ICSE)}.\hskip 1em plus 0.5em minus 0.4em\relax IEEE, 2023, pp. 1469--1481.

\bibitem{wei2023copiloting}
Y.~Wei, C.~S. Xia, and L.~Zhang, ``Copiloting the copilots: Fusing large language models with completion engines for automated program repair,'' in \emph{Proceedings of the 31st ACM Joint European Software Engineering Conference and Symposium on the Foundations of Software Engineering}, 2023, pp. 172--184.

\bibitem{zhang2024vuladvisor}
J.~Zhang, C.~Wang, A.~Li, W.~Wang, T.~Li, and Y.~Liu, ``Vuladvisor: Natural language suggestion generation for software vulnerability repair,'' in \emph{Proceedings of the 39th IEEE/ACM International Conference on Automated Software Engineering}, 2024, pp. 1932--1944.

\bibitem{nguyen2015graph}
A.~T. Nguyen and T.~N. Nguyen, ``Graph-based statistical language model for code,'' in \emph{2015 IEEE/ACM 37th IEEE International Conference on Software Engineering}, vol.~1.\hskip 1em plus 0.5em minus 0.4em\relax IEEE, 2015, pp. 858--868.

\bibitem{raychev2014code}
V.~Raychev, M.~Vechev, and E.~Yahav, ``Code completion with statistical language models,'' in \emph{Proceedings of the 35th ACM SIGPLAN conference on programming language design and implementation}, 2014, pp. 419--428.

\bibitem{ugare2024improving}
S.~Ugare, T.~Suresh, H.~Kang, S.~Misailovic, and G.~Singh, ``Improving llm code generation with grammar augmentation,'' \emph{arXiv preprint arXiv:2403.01632}, 2024.

\bibitem{guo2022unixcoder}
D.~Guo, S.~Lu, N.~Duan, Y.~Wang, M.~Zhou, and J.~Yin, ``Unixcoder: Unified cross-modal pre-training for code representation,'' \emph{arXiv preprint arXiv:2203.03850}, 2022.

\bibitem{copilot}
\BIBentryALTinterwordspacing
(2023) Github copilot. [Online]. Available: \url{https://github.com/features/copilot}
\BIBentrySTDinterwordspacing

\bibitem{Kula2018ESI}
R.~G. Kula, A.~Ouni, D.~M. German, and K.~Inoue, ``An empirical study on the impact of refactoring activities on evolving client-used apis,'' \emph{Inf. Softw. Technol.}, vol.~93, no.~C, pp. 186--199, 2018.

\bibitem{hu2023empirical}
M.~Hu and Y.~Zhang, ``An empirical study of the python/c api on evolution and bug patterns,'' \emph{Journal of Software: Evolution and Process}, vol.~35, no.~2, p. e2507, 2023.

\bibitem{sigsoft/Wang0LC20}
J.~Wang, L.~Li, K.~Liu, and H.~Cai, ``Exploring how deprecated python library apis are (not) handled,'' in \emph{Proceedings of the 28th acm joint meeting on european software engineering conference and symposium on the foundations of software engineering}, 2020, pp. 233--244.

\bibitem{APIlifecycle}
\BIBentryALTinterwordspacing
Api lifecycle stages. [Online]. Available: \url{https://developers.meetmarigold.com/engage/terms/versioning-deprecation/\#api-lifecycle-stages}
\BIBentrySTDinterwordspacing

\bibitem{PyTorch}
\BIBentryALTinterwordspacing
Pytorch: A python package that provides tensor computation and deep neural networks. [Online]. Available: \url{https://pytorch.org/}
\BIBentrySTDinterwordspacing

\bibitem{ijcai/ZanCYLKGWCL22}
D.~Zan, B.~Chen, D.~Yang, Z.~Lin, M.~Kim, B.~Guan, Y.~Wang, W.~Chen, and J.-G. Lou, ``Cert: continual pre-training on sketches for library-oriented code generation,'' \emph{arXiv preprint arXiv:2206.06888}, 2022.

\bibitem{corr/abs-2305-04032}
K.~Zhang, H.~Zhang, G.~Li, J.~Li, Z.~Li, and Z.~Jin, ``Toolcoder: Teach code generation models to use api search tools,'' \emph{arXiv preprint arXiv:2305.04032}, 2023.

\bibitem{xu2022systematic}
F.~F. Xu, U.~Alon, G.~Neubig, and V.~J. Hellendoorn, ``A systematic evaluation of large language models of code,'' in \emph{Proceedings of the 6th ACM SIGPLAN International Symposium on Machine Programming}, 2022, pp. 1--10.

\bibitem{liu2024your}
J.~Liu, C.~S. Xia, Y.~Wang, and L.~Zhang, ``Is your code generated by chatgpt really correct? rigorous evaluation of large language models for code generation,'' \emph{Advances in Neural Information Processing Systems}, vol.~36, 2024.

\bibitem{ciniselli2021empiricaltransformer}
M.~Ciniselli, N.~Cooper, L.~Pascarella, A.~Mastropaolo, E.~Aghajani, D.~Poshyvanyk, M.~Di~Penta, and G.~Bavota, ``An empirical study on the usage of transformer models for code completion,'' \emph{IEEE Transactions on Software Engineering}, vol.~48, no.~12, pp. 4818--4837, 2021.

\bibitem{ciniselli2021empiricalbert}
M.~Ciniselli, N.~Cooper, L.~Pascarella, D.~Poshyvanyk, M.~Di~Penta, and G.~Bavota, ``An empirical study on the usage of bert models for code completion,'' in \emph{2021 IEEE/ACM 18th International Conference on Mining Software Repositories (MSR)}.\hskip 1em plus 0.5em minus 0.4em\relax IEEE, 2021, pp. 108--119.

\bibitem{ding2023static}
H.~Ding, V.~Kumar, Y.~Tian, Z.~Wang, R.~Kwiatkowski, X.~Li, M.~K. Ramanathan, B.~Ray, P.~Bhatia, S.~Sengupta \emph{et~al.}, ``A static evaluation of code completion by large language models,'' \emph{arXiv preprint arXiv:2306.03203}, 2023.

\bibitem{izadi2024language}
M.~Izadi, J.~Katzy, T.~Van~Dam, M.~Otten, R.~M. Popescu, and A.~Van~Deursen, ``Language models for code completion: A practical evaluation,'' in \emph{Proceedings of the IEEE/ACM 46th International Conference on Software Engineering}, 2024, pp. 1--13.

\bibitem{liu2024exploring}
F.~Liu, Y.~Liu, L.~Shi, H.~Huang, R.~Wang, Z.~Yang, and L.~Zhang, ``Exploring and evaluating hallucinations in llm-powered code generation,'' \emph{arXiv preprint arXiv:2404.00971}, 2024.

\bibitem{zhang2024llm}
Z.~Zhang, Y.~Wang, C.~Wang, J.~Chen, and Z.~Zheng, ``Llm hallucinations in practical code generation: Phenomena, mechanism, and mitigation,'' \emph{arXiv preprint arXiv:2409.20550}, 2024.

\bibitem{sawant2018understanding}
A.~A. Sawant, M.~Aniche, A.~van Deursen, and A.~Bacchelli, ``Understanding developers' needs on deprecation as a language feature,'' in \emph{ICSE}, 2018, pp. 561--571.

\bibitem{sawant2018features}
A.~A. Sawant, G.~Huang, G.~Vilen, S.~Stojkovski, and A.~Bacchelli, ``Why are features deprecated? an investigation into the motivation behind deprecation,'' in \emph{ICSME}, 2018, pp. 13--24.

\bibitem{mirian2019web}
A.~Mirian, N.~Bhagat, C.~Sadowski, A.~P. Felt, S.~Savage, and G.~M. Voelker, ``Web feature deprecation: a case study for chrome,'' in \emph{ICSE-SEIP}, 2019, pp. 302--311.

\bibitem{sawant2019react}
A.~A. Sawant, R.~Robbes, and A.~Bacchelli, ``To react, or not to react: Patterns of reaction to api deprecation,'' \emph{Empirical Software Engineering}, vol.~24, no.~6, pp. 3824--3870, 2019.

\bibitem{robbes2012developers}
R.~Robbes, M.~Lungu, and D.~R{\"o}thlisberger, ``How do developers react to api deprecation? the case of a smalltalk ecosystem,'' in \emph{FSE}, 2012, pp. 1--11.

\bibitem{Linares-Vasquez2013ACF}
M.~Linares-V\'{a}squez, G.~Bavota, C.~Bernal-C\'{a}rdenas, M.~Di~Penta, R.~Oliveto, and D.~Poshyvanyk, ``Api change and fault proneness: A threat to the success of android apps,'' in \emph{ESEC/FSE}, 2013, pp. 477--487.

\bibitem{wang2020empirical}
Y.~Wang, B.~Chen, K.~Huang, B.~Shi, C.~Xu, X.~Peng, Y.~Wu, and Y.~Liu, ``An empirical study of usages, updates and risks of third-party libraries in java projects,'' in \emph{2020 IEEE International Conference on Software Maintenance and Evolution (ICSME)}.\hskip 1em plus 0.5em minus 0.4em\relax IEEE, 2020, pp. 35--45.

\bibitem{McDonnell2013ESA}
T.~McDonnell, B.~Ray, and M.~Kim, ``An empirical study of api stability and adoption in the android ecosystem,'' in \emph{ICSM}, 2013, pp. 70--79.

\bibitem{hora2015developers}
A.~Hora, R.~Robbes, N.~Anquetil, A.~Etien, S.~Ducasse, and M.~T. Valente, ``How do developers react to api evolution? the pharo ecosystem case,'' in \emph{ICSME}, 2015, pp. 251--260.

\bibitem{sawant2016reaction}
A.~A. Sawant, R.~Robbes, and A.~Bacchelli, ``On the reaction to deprecation of 25,357 clients of 4+ 1 popular java apis,'' in \emph{ICSME}, 2016, pp. 400--410.

\bibitem{Balaban2005RSC}
I.~Balaban, F.~Tip, and R.~Fuhrer, ``Refactoring support for class library migration,'' in \emph{OOPSLA}, 2005, pp. 265--279.

\bibitem{henkel2005catchup}
J.~Henkel and A.~Diwan, ``Catchup! capturing and replaying refactorings to support api evolution,'' in \emph{ICSE}, 2005, pp. 274--283.

\bibitem{xing2007api}
Z.~Xing and E.~Stroulia, ``Api-evolution support with diff-catchup,'' \emph{IEEE Transactions on Software Engineering}, vol.~33, no.~12, pp. 818--836, 2007.

\bibitem{dagenais2009semdiff}
B.~Dagenais and M.~P. Robillard, ``Semdiff: Analysis and recommendation support for api evolution,'' in \emph{ICSE}, 2009, pp. 599--602.

\bibitem{schafer2008mining}
T.~Sch{\"a}fer, J.~Jonas, and M.~Mezini, ``Mining framework usage changes from instantiation code,'' in \emph{ICSE}, 2008, pp. 471--480.

\bibitem{godfrey2005using}
M.~W. Godfrey and L.~Zou, ``Using origin analysis to detect merging and splitting of source code entities,'' \emph{IEEE Transactions on Software Engineering}, vol.~31, no.~2, pp. 166--181, 2005.

\bibitem{wu2010aura}
W.~Wu, Y.-G. Gu{\'e}h{\'e}neuc, G.~Antoniol, and M.~Kim, ``Aura: a hybrid approach to identify framework evolution,'' in \emph{ICSE}, 2010, pp. 325--334.

\bibitem{Huang2021}
K.~Huang, B.~Chen, L.~Pan, S.~Wu, and X.~Peng, ``Repfinder: Finding replacements for missing apis in library update,'' in \emph{ASE}, 2021.

\bibitem{sallou2024breaking}
J.~Sallou, T.~Durieux, and A.~Panichella, ``Breaking the silence: the threats of using llms in software engineering,'' in \emph{Proceedings of the 2024 ACM/IEEE 44th International Conference on Software Engineering: New Ideas and Emerging Results}, 2024, pp. 102--106.

\bibitem{popularlanguage}
\BIBentryALTinterwordspacing
Most popular programming languages. [Online]. Available: \url{https://www.orientsoftware.com/blog/most-popular-programming-languages/}
\BIBentrySTDinterwordspacing

\bibitem{pytorchdoc}
\BIBentryALTinterwordspacing
(2023) Pytorch documentation 1.9.0. [Online]. Available: \url{https://pytorch.org/docs/1.9.0/}
\BIBentrySTDinterwordspacing

\bibitem{Sourcegraph}
\BIBentryALTinterwordspacing
Openai api reference. [Online]. Available: \url{https://sourcegraph.com/search}
\BIBentrySTDinterwordspacing

\bibitem{PEP-221}
\BIBentryALTinterwordspacing
Pep 221 – import as. [Online]. Available: \url{https://peps.python.org/pep-0221/}
\BIBentrySTDinterwordspacing

\bibitem{Llama2}
H.~Touvron, L.~Martin, K.~Stone, P.~Albert, A.~Almahairi, Y.~Babaei, N.~Bashlykov, S.~Batra, P.~Bhargava, S.~Bhosale, and Others, ``Llama 2: Open foundation and fine-tuned chat models,'' \emph{CoRR}, vol. abs/2307.09288, 2023.

\bibitem{GPT-3.5}
\BIBentryALTinterwordspacing
Gpt-3.5 turbo. [Online]. Available: \url{https://platform.openai.com/docs/models/gpt-3-5-turbo}
\BIBentrySTDinterwordspacing

\bibitem{HuggingFace}
\BIBentryALTinterwordspacing
Hugging face - host git-based models, datasets and spaces on the hugging face hub. [Online]. Available: \url{https://huggingface.co/models}
\BIBentrySTDinterwordspacing

\bibitem{OpenAIAPI}
\BIBentryALTinterwordspacing
Sourcegraph: Code search and an ai assistant with the context of the code graph. [Online]. Available: \url{https://platform.openai.com/docs/api-reference/chat}
\BIBentrySTDinterwordspacing

\bibitem{COT}
J.~Wei, X.~Wang, D.~Schuurmans, M.~Bosma, B.~Ichter, F.~Xia, E.~H. Chi, Q.~V. Le, and D.~Zhou, ``Chain-of-thought prompting elicits reasoning in large language models,'' S.~Koyejo, S.~Mohamed, A.~Agarwal, D.~Belgrave, K.~Cho, and A.~Oh, Eds., 2022.

\bibitem{ICL}
A.~K. Lampinen, I.~Dasgupta, S.~C. Chan, K.~Matthewson, M.~H. Tessler, A.~Creswell, J.~L. McClelland, J.~X. Wang, and F.~Hill, ``Can language models learn from explanations in context?'' \emph{arXiv preprint arXiv:2204.02329}, 2022.

\bibitem{xie2023data}
S.~M. Xie, S.~Santurkar, T.~Ma, and P.~S. Liang, ``Data selection for language models via importance resampling,'' \emph{Advances in Neural Information Processing Systems}, vol.~36, pp. 34\,201--34\,227, 2023.

\bibitem{nijkamp2023codegen2}
E.~Nijkamp, H.~Hayashi, C.~Xiong, S.~Savarese, and Y.~Zhou, ``Codegen2: Lessons for training llms on programming and natural languages,'' \emph{arXiv preprint arXiv:2305.02309}, 2023.

\bibitem{LLM-KB}
F.~Petroni, T.~Rockt{\"a}schel, P.~Lewis, A.~Bakhtin, Y.~Wu, A.~H. Miller, and S.~Riedel, ``Language models as knowledge bases?'' \emph{arXiv preprint arXiv:1909.01066}, 2019.

\bibitem{LLM-KB-Revisiting}
B.~Cao, H.~Lin, X.~Han, L.~Sun, L.~Yan, M.~Liao, T.~Xue, and J.~Xu, ``Knowledgeable or educated guess? revisiting language models as knowledge bases,'' \emph{arXiv preprint arXiv:2106.09231}, 2021.

\bibitem{bengio2000neural}
Y.~Bengio, R.~Ducharme, and P.~Vincent, ``A neural probabilistic language model,'' \emph{Advances in neural information processing systems}, vol.~13, 2000.

\bibitem{Imbalance1}
J.~M. Johnson and T.~M. Khoshgoftaar, ``Survey on deep learning with class imbalance,'' \emph{J. Big Data}, vol.~6, p.~27, 2019.

\bibitem{Imbalance2}
X.~Liu, J.~Wu, and Z.~Zhou, ``Exploratory undersampling for class-imbalance learning,'' \emph{{IEEE} Trans. Syst. Man Cybern. Part {B}}, vol.~39, no.~2, pp. 539--550, 2009.

\bibitem{EditSim}
A.~Svyatkovskiy, S.~K. Deng, S.~Fu, and N.~Sundaresan, ``Intellicode compose: Code generation using transformer,'' in \emph{Proceedings of the 28th ACM joint meeting on European software engineering conference and symposium on the foundations of software engineering}, 2020, pp. 1433--1443.

\bibitem{Naturalness}
A.~Hindle, E.~T. Barr, M.~Gabel, Z.~Su, and P.~Devanbu, ``On the naturalness of software,'' \emph{Communications of the ACM}, vol.~59, no.~5, pp. 122--131, 2016.

\bibitem{PromptSensitivity1}
T.~Gao, A.~Fisch, and D.~Chen, ``Making pre-trained language models better few-shot learners,'' \emph{arXiv preprint arXiv:2012.15723}, 2020.

\bibitem{PromptSensitivity2}
Z.~Jiang, F.~F. Xu, J.~Araki, and G.~Neubig, ``How can we know what language models know?'' \emph{Transactions of the Association for Computational Linguistics}, vol.~8, pp. 423--438, 2020.

\bibitem{Mitchell22Memory}
E.~Mitchell, C.~Lin, A.~Bosselut, C.~D. Manning, and C.~Finn, ``Memory-based model editing at scale,'' in \emph{International Conference on Machine Learning}.\hskip 1em plus 0.5em minus 0.4em\relax PMLR, 2022, pp. 15\,817--15\,831.

\bibitem{Murty22Fixing}
S.~Murty, C.~D. Manning, S.~Lundberg, and M.~T. Ribeiro, ``Fixing model bugs with natural language patches,'' \emph{arXiv preprint arXiv:2211.03318}, 2022.

\bibitem{Madaan22Memory}
A.~Madaan, N.~Tandon, P.~Clark, and Y.~Yang, ``Memory-assisted prompt editing to improve gpt-3 after deployment,'' \emph{arXiv preprint arXiv:2201.06009}, 2022.

\bibitem{Meng22Locating}
K.~Meng, D.~Bau, A.~Andonian, and Y.~Belinkov, ``Locating and editing factual associations in gpt,'' \emph{Advances in Neural Information Processing Systems}, vol.~35, pp. 17\,359--17\,372, 2022.

\bibitem{Meng23Editing}
K.~Meng, A.~S. Sharma, A.~Andonian, Y.~Belinkov, and D.~Bau, ``Mass-editing memory in a transformer,'' \emph{arXiv preprint arXiv:2210.07229}, 2022.

\bibitem{Li24PMET}
X.~Li, S.~Li, S.~Song, J.~Yang, J.~Ma, and J.~Yu, ``Pmet: Precise model editing in a transformer,'' in \emph{Proceedings of the AAAI Conference on Artificial Intelligence}, vol.~38, no.~17, 2024, pp. 18\,564--18\,572.

\bibitem{Cao21Editing}
N.~De~Cao, W.~Aziz, and I.~Titov, ``Editing factual knowledge in language models,'' \emph{arXiv preprint arXiv:2104.08164}, 2021.

\bibitem{Mitchell22Fast}
E.~Mitchell, C.~Lin, A.~Bosselut, C.~Finn, and C.~D. Manning, ``Fast model editing at scale,'' \emph{arXiv preprint arXiv:2110.11309}, 2021.

\bibitem{Lewis20RAG}
P.~Lewis, E.~Perez, A.~Piktus, F.~Petroni, V.~Karpukhin, N.~Goyal, H.~K{\"u}ttler, M.~Lewis, W.-t. Yih, T.~Rockt{\"a}schel \emph{et~al.}, ``Retrieval-augmented generation for knowledge-intensive nlp tasks,'' \emph{Advances in Neural Information Processing Systems}, vol.~33, pp. 9459--9474, 2020.

\bibitem{gronauer2022multi}
S.~Gronauer and K.~Diepold, ``Multi-agent deep reinforcement learning: a survey,'' \emph{Artificial Intelligence Review}, vol.~55, no.~2, pp. 895--943, 2022.

\bibitem{talebirad2023multi}
Y.~Talebirad and A.~Nadiri, ``Multi-agent collaboration: Harnessing the power of intelligent llm agents,'' \emph{arXiv preprint arXiv:2306.03314}, 2023.

\bibitem{ellis2024smacv2}
B.~Ellis, J.~Cook, S.~Moalla, M.~Samvelyan, M.~Sun, A.~Mahajan, J.~Foerster, and S.~Whiteson, ``Smacv2: An improved benchmark for cooperative multi-agent reinforcement learning,'' \emph{Advances in Neural Information Processing Systems}, vol.~36, 2024.

\bibitem{wang2024towards}
C.~Wang, Z.~Chen, T.~Li, Y.~Zhao, and Y.~Liu, ``Towards trustworthy llms for code: A data-centric synergistic auditing framework,'' \emph{arXiv preprint arXiv:2410.09048}, 2024.

\bibitem{replication}
\BIBentryALTinterwordspacing
(2024) Replication package. [Online]. Available: \url{https://anonymous.4open.science/r/Replication-Deprecated-API-Usage-in-LLM-based-Code-Completion/README.md}
\BIBentrySTDinterwordspacing

\end{thebibliography}

\end{document}